\documentclass[aps, twocolumn, letterpaper, superscriptaddress, nofootinbib]{revtex4}

\usepackage{amsmath}
\usepackage{amssymb}
\usepackage{mathrsfs}
\usepackage{xspace}
\usepackage{color}
\usepackage{braket}
\usepackage{bbold}
\usepackage{graphicx}

\usepackage{ifpdf}
\ifpdf
\fi

\newcommand{\eq}[1]{(\ref{#1})}
\newcommand{\Eq}[1]{Eq.~(\ref{#1})}
\newcommand{\Eqs}[1]{Eqs.~(\ref{#1})}
\newcommand{\Fig}[1]{Fig.~\ref{#1}}
\newcommand{\Sec}[1]{Sec.~\ref{#1}}
\newcommand{\Ref}[1]{Ref.~\cite{#1}}
\newcommand{\Refs}[1]{Refs.~\cite{#1}}
\newcommand{\App}[1]{Appendix~\ref{#1}}

\newcommand{\eg}{{e.g.,\/}\xspace}
\newcommand{\ie}{{i.e.,\/}\xspace}
\newcommand{\etal}{{\it et~al.\/}\xspace}

\newcommand{\pd}{\partial}
\newcommand{\del}{\vec{\nabla}}
\newcommand{\mc}[1]{\mathcal{#1}}
\newcommand{\mcc}[1]{\mathfrak{#1}}
\newcommand{\msf}[1]{\mathsf{#1}}
\newcommand{\mcu}[1]{\mathscr{#1}}
\renewcommand{\vec}[1]{{\boldsymbol{\rm #1}}}
\newcommand{\favr}[1]{\langle #1 \rangle}

\newcommand{\ee}{\mathrm{e}}
\newcommand{\ii}{\mathrm{i}}
\newcommand{\dd}{\mathrm{d}}

\newcommand{\total}[1]{#1}
\newcommand{\aver}[1]{\bar{#1}}

\newcommand{\TT}{\mcu{T}}
\newcommand{\sech}{\mathrm{sech}}

\begin{document}

\title{Average nonlinear dynamics of particles in gravitational pulses: effective Hamiltonian, secular acceleration, and gravitational susceptibility}

\author{Deepen Garg}
\affiliation{Department of Astrophysical Sciences, Princeton University, Princeton, New Jersey 08544, USA}
\author{I. Y. Dodin}
\affiliation{Department of Astrophysical Sciences, Princeton University, Princeton, New Jersey 08544, USA}
\affiliation{Princeton Plasma Physics Laboratory, Princeton, NJ 08543, USA}

\date{\today}

\begin{abstract}
Particles interacting with a prescribed quasimonochromatic gravitational wave (GW) exhibit secular (average) nonlinear dynamics that can be described by Hamilton's equations. We derive the Hamiltonian of this ``ponderomotive'' dynamics to the second order in the GW amplitude for a general background metric. For the special case of vacuum GWs, we show that our Hamiltonian is equivalent to that of a free particle in an effective metric, which we calculate explicitly. We also show that already a linear plane GW pulse displaces a particle from its unperturbed trajectory by a finite distance that is independent of the GW phase and proportional to the integral of the pulse intensity. We calculate the particle displacement analytically and show that our result is in agreement with numerical simulations. We also show how the Hamiltonian of the \textit{nonlinear} averaged dynamics naturally leads to the concept of the \textit{linear} gravitational susceptibility of a particle gas with an arbitrary phase-space distribution. We calculate this susceptibility explicitly to apply it, in a follow-up paper, toward studying self-consistent GWs in inhomogeneous media within the geometrical-optics approximation.
\end{abstract}

\maketitle

\section{Introduction}

Recent detection of gravitational waves (GWs) \cite{ref:abbott16a, ref:abbott16b, ref:abbott17a, ref:abbott17b, ref:abbott17c, ref:abbott17d, ref:abbott19, ref:abbott20a, tex:abbott20b} is strengthening the interest of the physics community in GW--matter interactions. Linear effects of GWs have long been studied in literature \cite{ref:zeldovich74, ref:braginskii85, ref:braginsky87}, particularly in the context of GW dispersion in gases and plasmas \cite{ref:chesters73, ref:asseo76, ref:macedo83, ref:servin01, ref:moortgat03, ref:forsberg10a, ref:barta18}. Some authors have also explored the associated nonlinear phenomena, such as nonlinear memory effects \cite{ref:blanchet92, ref:christodoulou91, ref:thorne92, ref:wiseman91, ref:zhang18, ref:zhang17, ref:flanagan19}, the contribution of the GW tail from backscattering off the background curvature \cite{ref:blanchet92, ref:wiseman93}, and certain GW--plasma interactions \cite{ref:ignatev97, ref:brodin99, ref:brodin00,  ref:brodin00b, ref:papadopolous01, ref:brodin01, ref:balakin03c, ref:servin03a, ref:kallberg04, ref:vlahos04, ref:brodin05, ref:duez05a, ref:isliker06, ref:forsberg06, ref:farris08, ref:forsberg08, ref:brodin10b}. However, there remains another fundamental nonlinear effect, the ``ponderomotive'' effect, that is well-known for electromagnetic interactions \cite{ref:gaponov58, ref:motz67, ref:cary81, my:mneg, my:qdirpond} but has not yet received due attention in GW research. Like the aforementioned memory effects that have been known, the ponderomotive effect is hereditary, \ie depends on the whole GW-intensity profile. But unlike the known memory effects, the ponderomotive effect is determined by the particle-motion equations (not the Einstein equations), so it can be produced even by linear GWs propagating in flat background spacetime.

The essence of the ponderomotive effect by GWs is as follows. Since the particle motion equations in a given metric are nonlinear, a prescribed GW generally induces not just quiver but also secular (average) nonlinear dynamics, regardless of whether the wave itself is linear or not. This nonlinear dynamics of particles is generally too complicated to study analytically; but it can be made tractable for quasimonochromatic GWs. In this case, the particle average motion can be described by relatively simple Hamilton's equations, with a Hamiltonian that depends on the GW envelope and \textit{not} on the GW phase. To the lowest order, the GW contribution to this Hamiltonian is of the second order in the wave amplitude. The resulting perturbations to the particle trajectories can be significant near sources of gravitational radiation, where the metric oscillations are substantial. These perturbations can also be important when particles are exposed to GWs long enough, since the ponderomotive effect is phase-independent and cumulative (see below). Furthermore, the ponderomotive effect is inherently related to the \textit{linear} susceptibility of matter with respect to GWs. The corresponding statement for electromagnetic interactions is known as the $K$-$\chi$ theorem in plasma theory \cite{ref:cary77, ref:kaufman87} and has also been extended to more general Hamiltonian systems \cite{ref:kentwell87b, my:kchi, my:lens, my:qponder}. Hence, calculating the ponderomotive effect readily yields not just nonlinear forces on particles (which may or may not be significant in practice) but also linear dispersive properties of GWs in gases and plasmas. In this sense, the ponderomotive effect matters even in linear theory. 

Here, we calculate the ponderomotive effect by weak GWs on neutral particles in the general case, \ie when the GW envelope, wavevector, polarization, and background metric are arbitrary smooth functions of spacetime coordinates. Such general calculations are not easy to do by directly averaging the particle-motion equations, so we invoke variational methods that were recently developed within plasma theory for electromagnetic interactions \cite{my:itervar, my:bgk, my:acti, my:sharm}. We derive the Hamiltonian of the particle ponderomotive dynamics to the second order in the GW amplitude. For the special case of vacuum GWs, we show that our Hamiltonian is equivalent to that of a free particle in an effective metric, which we calculate explicitly. We also show that already a linear plane GW pulse displaces a particle from its unperturbed trajectory by a finite distance that is independent of the GW phase and proportional to the integral of the pulse intensity. In this sense, the ponderomotive effect is cumulative. We calculate the particle displacement analytically and show that our result is in agreement with numerical simulations. We also show how our general Hamiltonian yields the linear gravitational susceptibility of a particle gas with an arbitrary phase-space distribution. We calculate this susceptibility explicitly to apply it, in a follow-up paper, toward studying self-consistent GWs in inhomogeneous media within the geometrical-optics approximation.

Our paper is organized as follows. In \Sec{sec:pmotion}, we discuss the well-known equations of the particle motion in a prescribed metric, which we use later on. In \Sec{sec:basic}, we introduce the so-called oscillation-center formalism, which we build upon, by analogy with how this is done for electromagnetic interactions in plasma theory. In \Sec{sec:point}, we calculate the ponderomotive Hamiltonian and the ensuing equations of the average motion of a point particle. In \Sec{sec:fieldth}, we present an alternative derivation of the same ponderomotive Hamiltonian by treating particles as semiclassical quantum waves. We also apply these results to derive the gravitational susceptibility of a neutral gas. In \Sec{sec:vacuum}, we discuss the particle motion in a linear vacuum GW pulse as an example, and we derive the total displacement of a particle under the influence of such a pulse. In \Sec{sec:num}, we present test-particle simulations, which show good agreement with our analytic theory. In \Sec{sec:conclusions}, we summarize our main results. Supplementary calculations are given in appendices. In particular, \App{app:wave} details the derivation of a general theorem used in \Sec{sec:fieldoc}, and \App{app:chi} provides the derivation of an alternative form of the gravitational susceptibility introduced in \Sec{sec:polar}.

\section{Particle motion equations}
\label{sec:pmotion}

\subsection{Basic equations}
\label{sec:constraint}

Let us start with reviewing the known equations of the particle motion in a prescribed spacetime metric $g_{\alpha\beta}(x)$. We assume units such that the speed of light equals one ($c = 1$), and the metric signature is assumed to be $(-+++)$. Then, the action $S$ of a particle traveling between two fixed spacetime locations $x_{1,2} \doteq x(\tau_{1,2})$ is given by
\begin{gather}\label{eq:S}
S = - m\int_{\tau_1}^{\tau_2} \sqrt{-g_{\alpha\beta}u^\alpha u^\beta} \, \dd\tau.
\end{gather}
Here, the symbol $\doteq$ denotes definitions, $m$ is the particle mass, $u^\alpha \doteq \dd x^\alpha/\dd\tau$ is the particle four-velocity, $u_\alpha \doteq g_{\alpha\beta} u^\beta$, and the proper time $\tau$ is defined such that
\begin{gather}\label{eq:r}
u_{\alpha}u^{\alpha} = - 1.
\end{gather}
Equation \eq{eq:r} serves as a constraint on the variational principle that governs the particle motion. Deriving the motion equations rigorously for a constrained action can be a subtle issue. However, we can sidestep this issue by rewriting \Eq{eq:S} as an unconstrained action of the form
\begin{gather}
S=-m \int_{\tau_1}^{\tau_2} \, \dd\tau,
\label{eq:Stau}
\end{gather}
with \(\dd\tau = \sqrt{-g_{\alpha\beta}\dd x^\alpha \dd x^\beta}\). Since this $S$ is not quite of the usual form $\int \mcu{L}(x, \dd x/\dd\sigma)\,\dd \sigma$, the resulting motion equations are not quite the standard Euler--Lagrange equations either. However, these equations still can be derived straightforwardly. Below, we describe two known approaches to this problem in detail, because we will need to refer to details of these approaches in later sections.

\subsection{Covariant equations of motion}
 
One way to derive the particle-motion equations from \Eq{eq:Stau} is to proceed as follows \cite{book:landau2}. Consider a variation $x^\mu \to x^\mu + \delta x^\mu$ such that 
\begin{gather}\label{eq:bc}
\delta x^\mu(\tau_1) = \delta x^\mu(\tau_2) = 0.
\end{gather}
Then the variation of $S$ given by \Eq{eq:Stau} can be written~as 
\begin{align}
\delta S
& =-m\int_{\tau_1}^{\tau_2} \delta \, \dd\tau
\notag\\
& =m \int_{\tau_1}^{\tau_2} \frac{1}{2\dd\tau}\,\delta 
\left(g_{\alpha \beta}\,\dd x^{\alpha}\,\dd x^{\beta}\right)
\notag\\
& =m\int_{\tau_1}^{\tau_2} \frac{1}{2\dd\tau}\left(
\frac{\pd g_{\alpha \beta}}{\pd x^{\mu}}\,\delta x^{\mu}\,\dd x^{\alpha}\,\dd x^{\beta}
+2g_{\alpha \beta}\,\dd x^{\alpha}\,\dd \delta x^{\beta}
\right)
\notag\\
& =m\int_{\tau_1}^{\tau_2} \left(
\frac{1}{2}\frac{\pd g_{\alpha \beta}}{\pd x^{\mu}}\,u^{\alpha}u^{\beta}
-\frac{\dd u_{\mu}}{\dd\tau}
\right)\delta x^{\mu}\,\dd\tau.
\label{eq:dS}
\end{align}
[Here, we have used symmetry of $g_{\alpha\beta}$; we have also integrated by parts to obtain the last equality and used \Eq{eq:bc} to eliminate the boundary term.] Then, the requirement that $\delta S = 0$ for all $\delta x^\mu$ leads to the ``geodesic equation'':
\begin{gather}\label{eq:eu1}
\frac{\dd u_{\mu}}{\dd\tau}
=\frac{1}{2}\frac{\pd g_{\alpha \beta}}{\pd x^{\mu}}\,u^{\alpha}u^{\beta}.
\end{gather}

Equations \eq{eq:eu1} can be viewed as the Euler--Lagrange equations corresponding to the Lagrangian 
\begin{gather}\label{eq:Lxp}
L(x, u) = \frac{m}{2}\,[g_{\alpha\beta}(x) u^\alpha u^\beta - 1].
\end{gather}
The second term is constant and could be omitted, but we have introduced it to keep $L = - m$ on solutions [due to \Eq{eq:r}; this is consistent with \Eq{eq:Stau}] and to emphasize parallels with the calculations in the later sections. Let us also introduce the corresponding canonical momentum
\begin{gather}\label{eq:pa}
p_\alpha \doteq \frac{\pd L}{\pd u^\alpha} = m u_\alpha
\end{gather}
and the Hamiltonian $H \doteq p_\alpha u^\alpha - L$, or
\begin{gather}\label{eq:H}
H(x, p)=\frac{1}{2m}\,[g^{\alpha\beta}(x)p_\alpha p_\beta + m^2],
\end{gather}
where \(g^{\alpha\beta}\) is the inverse of the metric, \(g^{\alpha\mu} g_{\mu\beta}= \delta^\alpha_\beta\). (In later sections, we show how this Hamiltonian emerges more naturally from first principles.) The corresponding Hamilton's equations, equivalent to \Eq{eq:eu1}, are
\begin{gather}\label{eq:orays}
\frac{\dd x^\alpha}{\dd \tau} = \frac{\pd H}{\pd p_\alpha},
\quad
\frac{\dd p_\alpha}{\dd \tau} 
= - \frac{\pd H}{\pd x^\alpha},
\end{gather}
or explicitly,
\begin{gather}\label{eq:oorays}
\frac{\dd x^\alpha}{\dd \tau} = \frac{g^{\alpha\beta}p_\beta}{m},
\quad
\frac{\dd p_\alpha}{\dd \tau} 
= -\frac{1}{2m}\frac{\pd g^{\mu\nu}}{\pd x^\alpha}\,p_\mu p_\nu.
\end{gather}

\subsection{Non-covariant equations of motion}
\label{sec:noncovar0}

Another way to avoid dealing with the constraint \eq{eq:r} is to give up covariance of the motion equations and consider only the spatial dynamics instead \cite{ref:cognola86, foot:pmotion}. Let us use \Eq{eq:r} to express $u^0$ as a function of $t \doteq x^0$, $x^a$, and
\begin{gather}
v^a \doteq \frac{\dd x^a}{\dd t} = \frac{u^a}{u^0}.
\end{gather}
Specifically, $u^0 = \msf{u}^0(t, \vec{x}, \vec{v})$, where
\begin{gather}\label{eq:tgam}
(\msf{u}^0)^{-1} = \sqrt{-g_{ab}v^a v^b - 2g_{0b}v^b - g_{00}}.
\end{gather}
(Roman indices span from~1 to~3, unlike Greek indices, which span from~0 to~3. We also use bold font to denote three-dimensional spatial variables in the index-free form.) From \Eq{eq:oorays}, one has $\dd t/\dd \tau = u^0$, so $S$ can be written as a functional of only the spatial variables, $S = \int L_t\,\dd t$, where $L_t = - m/\msf{u}^0$. In this representation, the action is unconstrained, so the motion equations are the usual Euler--Lagrange equations,
\begin{gather}
\frac{\dd}{\dd t} \left(\frac{\pd L_t}{\pd v^a}\right) = \frac{\pd L_t}{\pd x^a}.
\end{gather}

Let us also introduce the corresponding Hamiltonian formulation. The canonical momenta are defined as $p_a = \smash{\pd L_t/\pd v^a}$,~so $p_a = m \smash{\msf{u}^0}(g_{a0} + \smash{g_{ab}v^b})$, or equivalently,
\begin{gather}\label{eq:cpi0}
p_a = m \left(g_{a0}\,\frac{\dd x^0}{\dd\tau} + g_{ab}\,\frac{\dd x^b}{\dd\tau}\right)
 = m g_{a\mu}u^\mu,
\end{gather}
where we used $\msf{u}^0 = \dd t/\dd \tau$. Therefore, these momenta are the same as the corresponding spatial components of the four-vector canonical momenta \eq{eq:pa}. Let us also consider $p^0 = m u^0$ and $p_0 = m g_{0\mu}u^\mu$ as functions of $(t, \vec{x}, \vec{p})$ and denote them as $\msf{p}^0(t, \vec{x}, \vec{p})$ and $\msf{p}_0(t, \vec{x}, \vec{p})$ respectively, 
\begin{gather}
\msf{p}^0 = m\msf{u}^0,
\quad \msf{p}_0 = m\msf{u}_0,
\end{gather}
where the latter satisfies
\begin{gather}
H(t, \vec{x}, \msf{p}_0(t, \vec{x}, \vec{p}), \vec{p}) = 0.
\label{eq:hamjac1}
\end{gather}
Using \Eq{eq:r}, we can find the explicit expressions for $\msf{p}^0(t, \vec{x}, \vec{p})$ and $\msf{p}_0(t, \vec{x}, \vec{p})$. In order to proceed, consider
\begin{equation}\label{eq:tildedh}
\sigma^{ab} \doteq g^{ab} - g^{a0}g^{b0}/g^{00}.
\end{equation}
Then, one can show that \cite{foot:pmotion}
\begin{gather}
\msf{p}^0(t, \vec{x}, \vec{p}) = m\gamma \sqrt{-g^{00}},
\label{eq:psup0}
\\
\msf{p}_0(t, \vec{x}, \vec{p}) = - \frac{m \gamma}{\sqrt{-g^{00}}} - \frac{g^{0a}}{g^{00}}\,p_a,
\label{eq:psub00}
\end{gather}
where $\gamma \doteq \sqrt{1 + \sigma^{ab}p_a p_b/m^2}$. One can further find the Hamiltonian of the spatial motion $H_t \doteq p_a v^a - L_t$ to be 
\begin{gather}\label{eq:ham3d}
H_t(t, \vec{x}, \vec{p}) = - \msf{p}_0(t, \vec{x}, \vec{p}).
\end{gather}
The corresponding Hamilton's equations are
\begin{gather}\label{eq:canon3d}
\frac{\dd x^a}{\dd t} = \frac{\pd H_t}{\pd p_a}, 
\quad 
\frac{\dd p_a}{\dd t} = - \frac{\pd H_t}{\pd x^a}.
\end{gather}
As can be checked, these equations are in agreement with the covariant Hamilton's equations \eq{eq:oorays}.

\section{Particles in an oscillating metric: basic concepts}
\label{sec:basic}

\subsection{Metric model}
\label{eq:mmodel}

Let us suppose a metric in the form
\begin{gather}\label{eq:metric}
\total{g}_{\alpha\beta} = \aver{g}_{\alpha\beta} + h_{\alpha\beta}.
\end{gather}
Here, $\aver{g}_{\alpha\beta}= \aver{g}_{\alpha\beta}(\epsilon x)$ is a slow function of the spacetime coordinates \(x\) and \(h_{\alpha\beta}\) is a quasimonochromatic perturbation, \ie can be expressed as \(h_{\alpha\beta} = h_{\alpha\beta}[\epsilon x, \theta(x)]\), where \(\epsilon\) is a small parameter and the dependence on the scalar ``phase'' $\theta$ is $2\pi$-periodic. We also assume
\begin{gather}\label{eq:eta}
h_{\alpha\beta} \ll 1, 
\quad
\favr{h_{\alpha\beta}}_\theta = 0,
\end{gather}
where $\favr{\ldots}_\theta$ denotes average over $\theta$. Then, \(\aver{g}_{\alpha\beta}\) can be understood as the $\theta$-average part of the total metric,
\begin{gather}
\aver{g}_{\alpha\beta} = \favr{\total{g}_{\alpha\beta}}_\theta.
\end{gather}
We shall attribute such metric perturbation as a GW. Note that 
\begin{gather}\label{eq:kdef}
k_\alpha \doteq \pd_\alpha \theta = \nabla_\alpha \theta
\end{gather}
can be interpreted as the local wavevector and $\epsilon$ can be interpreted as the geometrical-optics (GO) parameter, which is roughly
\begin{gather}
\epsilon \sim \lambda/\ell \ll 1.
\end{gather}
Here, $\lambda$ is the characteristic wavelength (in spacetime) and~$\ell \sim [\min\lbrace \pd \aver{g}(\epsilon x), \pd h(\epsilon x, \theta), \pd \lambda(\epsilon x) \rbrace]^{-1}$ is the characteristic inhomogeneity scale (in spacetime) of the background metric, GW envelope, and GW wavelength.

Note that the GW is \textit{not} assumed linear. The quasiperiodic functions $\smash{h_{\alpha\beta}}$ may contain multiple harmonics, and any secular nonlinearity can be absorbed in the background metric $\smash{\bar{g}_{\alpha\beta}}$. Hence, the latter can be responsible for various nonlinear memory effects additional to the ponderomotive effect derived in this paper. But for our purposes, $\smash{\bar{g}_{\alpha\beta}}$ does not need to be specified, so those additional memory effects will not be articulated.

\subsection{Oscillation-center coordinates}

Let us consider the particle motion in the metric \eq{eq:metric}. We shall assume that a particle oscillates many times while traveling the distance $\ell$. [We shall also assume, to avoid introducing additional parameters, that the corresponding number of oscillations is $O(\epsilon^{-1})$.] Then, its motion is quasiperiodic in time, and one can use standard methods of plasma theory~\cite{ref:brizard09} to construct new ``oscillation-center'' (OC) coordinates $X$ in which the particle dynamics is non-oscillatory. This amounts to replacing the original particles with OCs, or ``dressed'' particles, that do not exhibit oscillations. Here, we adopt a less formal and perhaps more intuitive approach to  construct the same transformation to the leading order in~$\epsilon$. 

Let us start by introducing the local time average
\begin{gather}\label{eq:avr}
\favr{f}_t \doteq \frac{\int_{\Delta T} f\,\dd t}{\int_{\Delta T} \dd t},
\end{gather}
where $\Delta T$ is much larger than the oscillation period yet small enough such that the particle motion during this time remains approximately periodic. Then, the particle coordinates $x^\alpha$ can be separated into the slow OC coordinates $X^\alpha \doteq \favr{x^\alpha}_t$ and the quiver displacements $\widetilde{x}^\alpha(X, V)$ with zero time average: 
\begin{gather}\label{eq:x}
\total{x}^\alpha = X^\alpha + \widetilde{x}^\alpha,
\quad
\favr{\widetilde{x}^\alpha}_t = 0.
\end{gather}
Similarly, we introduce the OC velocities $V^\alpha$ as $\Delta X^\alpha/\Delta T$, where $\Delta T$ is used in the same way as in \Eq{eq:avr}. Then, one finds from \Eq{eq:avr} that $V^\alpha = \favr{v^\alpha}_t$. In particular, $V^0 = 1$. Also note that $V^\alpha$ can be understood as the derivatives of $X^\alpha$ with respect to the OC time $X^0 \equiv T$, \ie $V^\alpha = \dd X^\alpha/\dd T \equiv \Delta X/\Delta T$. Note that as introduced here, the ``infinitesimal'' OC displacements are well-defined only as averages over many oscillation cycles. (However, this limitation is waived in the more formal approach to the OC dynamics \cite{ref:brizard09}.)

Using \Eq{eq:avr} and $\dd t = u^0 \dd \tau$, we find that the average of any quasiperiodic function $f$ over $t$ and the corresponding local average over $\tau$ satisfy
\begin{gather}
\favr{f}_t
= \frac{\int f u^0\dd \tau}{\int u^0\dd \tau}
= \frac{\favr{f u^0}_\tau}{\favr{u^0}_\tau}.
\end{gather}
Hence, the OC velocities can be expressed as follows:
\begin{gather}
V^\alpha = \favr{\dot{x}^\alpha}_t = \frac{\favr{u^\alpha}_\tau}{\favr{u^0}_\tau}.
\end{gather}
Introducing $U^\alpha \doteq \favr{u^\alpha}_\tau$, we get $V^\alpha = U^\alpha / U^0$, and
\begin{gather}\label{eq:u}
u^\alpha = U^\alpha + \widetilde{u}^\alpha, 
\quad 
\favr{\widetilde{u}^\alpha}_\tau = 0.
\end{gather}
Also, on an interval $\Delta \tau$ that includes multiple oscillations but is smaller than the characteristic scale of the OC motion, one has
\begin{gather}\label{eq:XU}
U^\alpha = \frac{\Delta X^\alpha}{\Delta \tau} = 
\frac{\dd X^\alpha}{\dd \tau}, 
\end{gather}
where, like in the case of $V^\alpha$, the ``infinitesimal'' OC displacements are understood as nonvanishing displacements averaged over many oscillations.

The \(\tau\)-average that enters the above formulas is connected with the \(\theta\)-average introduced in \Sec{eq:mmodel} via
\begin{gather}
\label{eq:tauaverage}
\favr{f}_\tau 
= \frac{\int_0^{2\pi} f\,\frac{\dd\tau}{\dd\theta}\,\dd\theta}%
{\int_0^{2\pi} \dd\theta}
=\frac{\favr{f \Omega^{-1}}_\theta}{\favr{\Omega^{-1}}_\theta},
\end{gather}
where $\Omega$ is the ``proper frequency'' given by
\begin{gather}\label{eq:Omega}
\Omega \doteq \frac{\dd\theta}{\dd\tau} 
= k_\alpha u^\alpha.
\end{gather}
Note that $\Omega$ can be also be expressed as
\begin{gather}
\Omega = \favr{\Omega}_\theta + \widetilde{\Omega}, 
\quad
\widetilde{\Omega} = O(h),
\quad 
\favr{\widetilde{\Omega}}_\theta = 0.
\end{gather}
Hence, $\Omega^{-1} = \favr{\Omega}_\theta^{-1} - \widetilde{\Omega}\favr{\Omega}_\theta^{-2} + O(h^2)$, so from \Eq{eq:tauaverage}, one obtains
\begin{gather}\label{eq:ftau2}
\favr{f}_\tau 
=\favr{f}_\theta - \frac{\favr{f\widetilde{\Omega}_\theta}_\theta}{\favr{\Omega}_\theta} + O(h^2),
\end{gather}
which yields $\favr{\Omega}_\tau - \favr{\Omega}_\theta =O(h^2)$. Since $\favr{\Omega}_\tau = k_\alpha U^\alpha$, this leads to the following formulas, which we use later:
\begin{gather}\label{eq:Omaver}
\favr{\Omega}_\theta = k_\alpha U^\alpha + O(h^2),
\quad
\widetilde{\Omega} = k_\alpha\widetilde{u}^\alpha + O(h^2).
\end{gather}

\subsection{Linear and nonlinear dynamics}
\label{sec:ponderbasic}

Using \Eq{eq:metric}, it is readily seen that \cite[Sec.~105]{book:landau2}
\begin{gather}
g^{\alpha\beta} = \bar{g}^{\alpha\beta}-h^{\alpha\beta}+{h^\alpha}_\gamma h^{\gamma\beta} + O(h^3),
\label{eq:hupper}
\end{gather}
where $h$ denotes the characteristic value of $h_{\alpha\beta}$ and $O(h^3)$ is henceforth neglected. Note that here and further, indices in $h_{\alpha\beta}$ are raised using the inverse of the background metric, \(\bar{g}^{\alpha\beta}\). Using $u^\alpha = g^{\alpha\beta}u_\beta$ and \Eqs{eq:oorays}, we find
\begin{gather}\label{eq:aux1}
\frac{\dd u^\alpha}{\dd \tau} = u_\beta\frac{\dd g^{\alpha\beta}}{\dd \tau}-\frac{1}{2}\,g^{\alpha\beta}\frac{\pd g^{\mu\nu}}{\pd x^\beta}u_\mu u_\nu.
\end{gather}
To the lowest order in $h$, one has from \Eq{eq:hupper} that $g^{\alpha\beta} \simeq \bar{g}^{\alpha\beta}-h^{\alpha\beta}$. Also,
\begin{gather}
\frac{\pd h^{\mu\nu}}{\pd x^\beta}
\simeq \frac{\dd h^{\mu\nu}}{\dd \theta}\,\frac{\pd\theta}{\pd x^\beta}
= \frac{\dd h^{\mu\nu}}{\dd \theta}\, k_\beta
\simeq \frac{\dd h^{\mu\nu}}{\dd \tau}\, \frac{k_\beta}{k_\lambda U^\lambda},
\end{gather}
where we have substituted \Eq{eq:Omega} and ignored $O(\epsilon)$ corrections. Hence, \Eq{eq:aux1} leads~to
\begin{gather}
\frac{\dd u^\alpha}{\dd \tau} \simeq -U_\beta\frac{\dd h^{\alpha\beta}}{\dd \tau}+\frac{g^{\alpha\beta}k_\beta}{2k_\lambda U^\lambda}\,U_\mu U_\nu\,\frac{\dd h^{\mu\nu}}{\dd \tau}.
\end{gather}
This can be readily integrated, yielding $U^\alpha \simeq \text{const}$ and
\begin{gather}
\widetilde{u}^\alpha \simeq -h^{\alpha\nu}U_\nu + h^{\mu\nu} \frac{k^\alpha U_\mu U_\nu}{2 k_\lambda U^\lambda},
\label{eq:ua}
\end{gather}
where, within the assumed accuracy, the indices are manipulated using the background metric.

Note that this result is only a linear approximation. If the second and higher orders in~$h$ are retained in the equation for $u^\alpha$, one finds that a particle experiences a nonvanishing average force from a rapidly oscillating GW, if the GW is inhomogeneous or propagates in an inhomogeneous background. In analogy with electromagnetic interactions, this effect can be understood as the \textit{average gravitational ponderomotive force}. Our goal is to calculate this force and to describe its effects on the particle motion by studying the OC, or secular, dynamics.

One way to derive OC equations is by directly time-averaging the equations for $(x, v)$, which can be obtained from \Eqs{eq:oorays}. However, this approach is cumbersome and not particularly instructive. More instructive is the average-Lagrangian approach, which yields a manifestly Hamiltonian form of the motion equation. (This approach is also used to describe the dynamics of plasma particles in intense electromagnetic waves; see \Ref{my:itervar} for an overview.) Below, we consider two versions of this approach. In \Sec{sec:point}, we present a ``point-particle'' calculation, which is more direct but less tractable. In \Sec{sec:fieldth}, we present a ``field-theoretical'' calculation, which is less straightforward but yields the same results more transparently and in a form advantageous for the applications discussed in \Sec{sec:polar}.

\section{Oscillation-center dynamics: point-particle approach}
\label{sec:point}

Let us express the action \eq{eq:S} as $S = \int L_\tau\,\dd\tau$ (the integration limits are henceforth omitted for brevity), where\footnote{As discussed in \Sec{sec:constraint}, the function $L_\tau$ is \textit{not} a Lagrangian. It is used here only as a means to calculate the value of $S$, which is the same in \Eqs{eq:S} and \eq{eq:Stau}. How to infer motion equations from this value will be discussed in Secs.~\ref{sec:covarOC1} and \ref{sec:covarOC2}.}
\begin{gather}\label{eq:Ltau}
L_\tau = -m\sqrt{-g_{\alpha \beta}(x)u^{\alpha}u^{\beta}}.
\end{gather}
After substituting \Eq{eq:x}, one can express $L_\tau$ as a sum of $\bar{L}_\tau \doteq \favr{L_\tau}_\tau$, which is a slow function of the OC variables, and $\smash{\widetilde{L}_\tau}$, whose local $\tau$-average over rapid oscillations is zero. Since $\widetilde{L}_\tau$ does not contribute to $S$ at large enough~$\tau$, one obtains 
\begin{gather}\label{eq:SbL}
S \simeq \int \bar{L}_\tau\,\dd \tau.
\end{gather}

\subsection{Average action}

To calculate $\bar{L}_\tau$, we proceed as follows. From \Eqs{eq:metric} and \eq{eq:u}, we have
\begin{multline}
L_\tau = - m \sqrt{-(\bar{g}_{\alpha\beta}+h_{\alpha\beta})(U^\alpha+\widetilde{u}^\alpha)( U^\beta+\widetilde{u}^\beta})
\\
= -m \sqrt{-\bar{g}_{\alpha\beta}U^\alpha U^\beta}\sqrt{1 + \varphi},
\end{multline}
where
\begin{gather}\notag
\varphi = \frac{
h_{\alpha\beta}U^\alpha U^\beta
+ 2\bar{g}_{\alpha\beta} \widetilde{u}^{(\alpha} U^{\beta)}
+ 2h_{\alpha\beta}\widetilde{u}^{(\alpha} U^{\beta)}
+ \bar{g}_{\alpha\beta}\widetilde{u}^\alpha\widetilde{u}^\beta
}{\bar{g}_{\alpha\beta}U^\alpha U^\beta}.
\end{gather}
Hence, to the second order in $h$,
\begin{gather}
\bar{L}_\tau = \bar{L}_\tau^{(0)}\left(
1 + \frac{\favr{\varphi}_\tau}{2} - \frac{\favr{\varphi^2}_\tau}{8}
\right),\\
\bar{L}_\tau^{(0)} \doteq -m \sqrt{-\bar{g}_{\alpha\beta}U^\alpha U^\beta}.
\end{gather}
Within the same accuracy,
\begin{multline}
\favr{\varphi}_\tau = - \bar{g}_{\alpha\beta}\favr{\widetilde{u}^\alpha\widetilde{u}^\beta}_\tau
-\favr{h_{\alpha\beta}\widetilde{u}^{\alpha}}_\tau U^{\beta}
\\-\favr{h_{\alpha\beta}\widetilde{u}^{\beta}}_\tau U^{\alpha}
-\favr{h_{\alpha\beta}}_\tau U^\alpha U^\beta,
\end{multline}
\begin{multline}
\favr{\varphi^2}_\tau = 4U_\alpha U_\beta \favr{\widetilde{u}^\alpha\widetilde{u}^\beta}_\tau
+ 4 U^\alpha U^\beta U_\gamma \favr{h_{\alpha\beta}\widetilde{u}^\gamma}_\tau
\\ + \favr{h_{\alpha\beta}h_{\gamma\delta}}_\tau U^\alpha U^\beta U^\gamma U^\delta,
\end{multline}
where we used that to the leading (zeroth) order in $h$, one has $\bar{g}_{\alpha\beta}U^\alpha U^\beta \simeq -1$. Hence, $\bar{L}_\tau = \smash{\bar{L}_\tau^{(0)}} + \smash{\bar{L}_\tau^{(2)}}$, where
\begin{align}
\frac{2\bar{L}_\tau^{(2)}}{m}
= & \,\favr{\widetilde{u}^\alpha\widetilde{u}^\beta}_\tau 
(\bar{g}_{\alpha\beta} + U_\alpha U_\beta)
\notag \\
& + \favr{h_{\alpha\beta}\widetilde{u}^\gamma}_\tau 
(\delta_\gamma^{\alpha} U^{\beta} + \delta_\gamma^{\beta} U^{\alpha} + U^\alpha U^\beta U_\gamma)
\notag \\
& + \frac{1}{4} \favr{h_{\alpha\beta}h_{\gamma\delta}}_\tau U^\alpha U^\beta U^\gamma U^\delta
\notag \\
& + \favr{h_{\alpha\beta}}_\tau U^\alpha U^\beta.
\label{eq:L2}
\end{align}
The terms in the first three angular brackets are already of order $h^2$, so averaging over $\tau$ can be replaced with averaging over $\theta$. Then, using \Eq{eq:ua}, we obtain
\begin{multline}
\favr{\widetilde{u}^\alpha\widetilde{u}^\beta}_\tau=\varepsilon_{\mu\nu\gamma\delta}U^\nu U^\delta
\bigg[
\bar{g}^{\alpha\mu}\bar{g}^{\beta\gamma}
+
\frac{U^\mu U^\gamma k^\alpha k^\beta}{4(k_\lambda U^\lambda)^2}
\\
-\frac{\bar{g}^{\mu\alpha}U^\gamma k^\beta}{k_\lambda U^\lambda}
-\frac{\bar{g}^{\gamma\beta}U^\mu k^\alpha}{k_\lambda U^\lambda}
\bigg],
\end{multline}
\begin{gather}
\favr{h_{\alpha\beta}\widetilde{u}^\gamma}_\tau=\varepsilon_{\alpha\beta\mu\nu}U^\nu\left(\frac{U^\mu k^\gamma}{2k_\lambda U^\lambda}-\bar{g}^{\mu\gamma}\right),
\label{eq:hua}
\\
\favr{h_{\alpha\beta}h_{\gamma\delta}}_\tau = \varepsilon_{\alpha\beta\gamma\delta},
\end{gather}
where \(\varepsilon_{\alpha\beta\gamma\delta}\) is given by 
\begin{gather}
\varepsilon_{\alpha\beta\gamma\delta}\doteq\favr{h_{\alpha\beta}h_{\gamma\delta}}_\theta.
\label{eq:varepsilon}
\end{gather}
Also, from \Eqs{eq:ftau2} and \eq{eq:Omaver}, one has
\begin{multline}
\favr{h_{\alpha\beta}}_\tau 
\simeq - \frac{\favr{h_{\alpha\beta}\widetilde{\Omega}}_\theta} {k_\lambda U^\lambda}
= -\frac{\favr{h_{\alpha\beta} k_\gamma \widetilde{u}^\gamma}_\theta}{k_\lambda U^\lambda}
\\
=\frac{\varepsilon_{\alpha\beta\mu\nu}U^\nu}{k_\lambda U^\lambda}
\left(k^\mu - \frac{U^\mu k_\gamma k^\gamma}{2k_\lambda U^\lambda}
\right),
\end{multline}
where we used \Eq{eq:hua} in the last step. Then, from \Eq{eq:L2}, one obtains
\begin{multline}
\bar{L}_\tau^{(2)}=
-\varepsilon_{\alpha\beta\gamma\delta}\,\frac{m}{2}\,\bigg[
\bar{g}^{\beta\gamma}U^\alpha U^\delta +
\frac{k_\mu k^\mu}{4(k_\lambda U^\lambda)^2}U^\alpha U^\beta U^\gamma U^\delta \\
-\frac{1}{k_\lambda U^\lambda} U^\alpha U^\beta k^\gamma U^\delta
\bigg],
\label{eq:Lb2}
\end{multline}
where we have used the symmetry of $\varepsilon_{\alpha\beta\gamma\delta}$ with respect to index permutations $\alpha \leftrightarrow \beta$, $\gamma \leftrightarrow \delta$, and $(\alpha, \beta) \leftrightarrow (\gamma, \delta)$. Finally, the OC action can be expressed as
\begin{gather}\label{eq:SLoc}
S \simeq \int (-m + \bar{L}_\tau^{(2)})\,\dd\bar{\tau},\\
\dd\bar{\tau} \doteq \sqrt{-\bar{g}_{\alpha\beta}\,\dd X^\alpha\, \dd X^\beta},
\end{gather}
where we have used \Eq{eq:XU} and ignored higher order terms.

\subsection{Covariant equations of motion}
\label{sec:covarOC1}

Using \Eq{eq:SLoc}, the OC motion equations are obtained as follows. First, note that
\begin{gather}\label{eq:dS2t}
\delta S = \underbrace{-m \int \delta\,\dd\bar{\tau}}_{\delta S_1}
+ \underbrace{\int \delta\bar{L}_\tau^{(2)}\,\dd\bar{\tau}}_{\delta S_2}
+ \underbrace{\int \bar{L}_\tau^{(2)}\,\delta\,\dd\bar{\tau}}_{\delta S_3}.
\end{gather}
The first integral in \Eq{eq:dS2t} is calculated as in \Eq{eq:dS},
\begin{gather}\label{eq:dS1}
\delta S_1 = m\int \left(
\frac{1}{2}\frac{\pd \bar{g}_{\alpha \beta}}{\pd X^{\mu}}\,U^{\alpha}U^{\beta}
-\frac{\dd U_{\mu}}{\dd\bar{\tau}}
\right)\delta X^{\mu}\,\dd\bar{\tau}.
\end{gather}
The second integral in \Eq{eq:dS2t} is as usual,
\begin{gather}\label{eq:dS2}
\delta S_2
= \int \bigg[
\frac{\pd \bar{L}_\tau^{(2)}}{\pd X^\mu} - \frac{\dd}{\dd \bar{\tau}}\bigg(
\frac{\pd \bar{L}_\tau^{(2)}}{\pd U^\mu}
\bigg)
\bigg]\delta X^\mu\,\dd\bar{\tau}.
\end{gather}
The third integral in \Eq{eq:dS2t} is [cf. \Eq{eq:dS}]
\begin{multline}\label{eq:dS3}
\delta S_3
=m\int \left(
\frac{1}{2}
\frac{\pd \bar{g}_{\alpha \beta}}{\pd X^{\mu}}\,U^{\alpha}U^{\beta}\,\delta X^{\mu}
+U_{\mu}\,\frac{\dd\delta X^{\mu}}{\dd\bar{\tau}}
\right)\bar{L}_\tau^{(2)}\,\dd\bar{\tau}
\\
= \delta S_4 - \int \frac{\dd\bar{L}_\tau^{(2)}}{\dd\bar{\tau}}\,U_\mu\,\delta X^{\mu}\,\dd\bar{\tau},
\end{multline}
where
\begin{gather}\label{eq:dS4}
\delta S_4 =
m\int \left(
\frac{1}{2}
\frac{\pd \bar{g}_{\alpha \beta}}{\pd X^{\mu}}\,U^{\alpha}U^{\beta}
- \frac{\dd U_{\mu}}{\dd\bar{\tau}}
\right)\bar{L}_\tau^{(2)}\,\delta X^{\mu}\,\dd\bar{\tau}.
\end{gather}
To the zeroth order in $h$, OCs travel along geodesics of the unperturbed metric. Thus, the expression in parenthesis in \Eq{eq:dS4} is $o(h^0)$ and $\smash{\bar{L}_\tau^{(2)} = O(h^2)}$. Therefore, $\delta S_4 = o(h^2)$ and will be neglected. Then, from $\delta S = 0$ and \Eqs{eq:dS1}--\eq{eq:dS3}, one obtains the following equation:
\begin{multline}
m\bigg(\frac{1}{2}\frac{\pd \bar{g}_{\alpha \beta}}{\pd X^{\mu}}\,U^{\alpha}U^{\beta}
-\frac{\dd U_{\mu}}{\dd\bar{\tau}}\bigg)
+ \frac{\pd \bar{L}_\tau^{(2)}}{\pd X^\mu} - \frac{\dd}{\dd \bar{\tau}}\bigg(
\frac{\pd \bar{L}_\tau^{(2)}}{\pd U^\mu}\bigg)
\\= \frac{\dd\bar{L}_\tau^{(2)}}{\dd\bar{\tau}}\,U_\mu.
\label{eq:nEL}
\end{multline}

Let us introduce the new time $\mc{T}$ via $\dd \mc{T}/\dd\bar{\tau} = 1 + \zeta$, where $\zeta = O(h^2)$ is yet to be defined. Then,
\begin{multline}\label{eq:ELC}
m\bigg(\frac{1}{2}\frac{\pd \bar{g}_{\alpha \beta}}{\pd X^{\mu}}\,W^{\alpha}W^{\beta}
-\frac{\dd W_{\mu}}{\dd\mc{T}}
\bigg)
+ \frac{\pd \bar{L}_\tau^{(2)}}{\pd X^\mu} 
\\- \frac{\dd}{\dd \mc{T}}\bigg(
\frac{\pd \bar{L}_\tau^{(2)}}{\pd W^\mu}\bigg)
= m\mc{C},
\end{multline}
where $W^\alpha \doteq \dd X^\alpha/\dd\mc{T}$ and
\begin{gather}\notag
\mc{C} \simeq
W_{\mu}\,\frac{\dd}{\dd\mc{T}}\,\bigg(\zeta + \frac{\bar{L}_\tau^{(2)}}{m}\bigg)
-2\zeta\left(
\frac{1}{2}\frac{\pd \bar{g}_{\alpha \beta}}{\pd X^{\mu}}\,W^{\alpha}W^{\beta}
- \frac{\dd W_{\mu}}{\dd\mc{T}}
\right)
.
\end{gather}
Like in \Eq{eq:dS4}, the expression in the second parenthesis is $o(h^0)$ and $\smash{\zeta = O(h^2)}$, so the second term is $o(h^2)$ and is, therefore, negligible. Then, adopting $
\smash{\zeta = -\bar{L}_\tau^{(2)}}/m$, or
\begin{gather}\label{eq:Ttau}
\dd \mc{T}/\dd\bar{\tau} = 1 -\bar{L}_\tau^{(2)}/m,
\end{gather}
allows one to neglect the whole $\mc{C}$. In this case, \Eq{eq:ELC} can be viewed as an Euler--Lagrange equation
\begin{gather}
\frac{\dd}{\dd \mc{T}}\left(
\frac{\pd \mc{L}}{\pd W^\mu}
\right) = \frac{\pd \mc{L}}{\pd X^\mu}
\end{gather}
that corresponds to the following Lagrangian~[cf.~\Eq{eq:Lxp}]:
\begin{gather}
\mc{L}(X, W) = \frac{m}{2}\,[\bar{g}_{\alpha \beta}(X) W^\alpha W^\beta- 1] 
+ \bar{L}_\tau^{(2)}(X, W).
\end{gather}

Let us also introduce the OC canonical momentum
\begin{gather}
P_\alpha \doteq \frac{\pd \mc{L}}{\pd W^\alpha} = 
m \bar{g}_{\alpha \beta}W^\beta + \frac{\pd \bar{L}_\tau^{(2)}}{\pd W^\alpha}
\end{gather}
and the OC Hamiltonian $\mc{H} \doteq P_\alpha W^\alpha - \mc{L}$. Since $\bar{L}_\tau^{(2)}$ is small, a general theorem \cite[Sec.~40]{book:landau1} yields that
\begin{gather}\label{eq:mcHc1}
\mc{H} = H^{(0)} + \mc{H}^{(2)}, \quad \mc{H}^{(2)} = - \bar{L}_\tau^{(2)}.
\end{gather}
to the first nonvanishing order in the perturbation. The function $H^{(0)}$ is the unperturbed Hamiltonian, \ie
\begin{gather}\label{eq:H0}
H^{(0)}(X, P) = \frac{1}{2m}\,[\bar{g}^{\alpha\beta}(X) P_\alpha P_\beta + m^2],
\end{gather}
and one can adopt the lowest-order approximation $W^\alpha \simeq \bar{g}^{\alpha \beta}P_\beta/m$ when evaluating $\bar{L}_\tau^{(2)}$. This leads to
\begin{multline}\label{eq:mcHc3}
\mc{H}^{(2)} = \frac{\varepsilon^{\alpha\beta\gamma\delta}}{2m}\bigg[
\bar{g}_{\beta\gamma}\TT_{\alpha\delta}
+ \frac{k_\mu k^\mu}{4(k^\lambda P_\lambda)^2}\,\TT_{\alpha\beta} \TT_{\gamma\delta} 
\\-\frac{\TT_{\alpha\beta}}{k^\lambda P_\lambda} k_\gamma P_\delta
\bigg],
\end{multline}
where we have introduced
\begin{gather}
\TT_{\alpha\beta} \doteq P_\alpha P_\beta.
\label{eq:TT}
\end{gather}
The OC motion equations corresponding to the Hamiltonian \eq{eq:mcHc1} are
\begin{gather}\label{eq:hameqs}
\frac{\dd X^\alpha}{\dd \mc{T}} 
= \frac{\pd \mc{H}(X, P)}{\pd P_\alpha},
\quad
\frac{\dd P_\alpha}{\dd \mc{T}} 
= -\frac{\pd \mc{H}(X, P)}{\pd X^\alpha}.
\end{gather}

\subsection{Non-covariant equations of motion}
\label{sec:covarOC2}

Let us also derive these equations in a non-covariant form that, in particular, will be useful in \Sec{sec:gravpolariz}. Using \Eq{eq:XU} for $U^0$, one can rewrite the OC action \eq{eq:SbL} as $S = \int \msf{L}\,\dd T$, with $\msf{L} \doteq \bar{L}_\tau/U^0$, and consider $S$ as a functional of $\vec{X}(T)$. Like in \Sec{sec:noncovar0}, the variational principle for the spatial dynamics is unconstrained, so $\msf{L}$ can be understood as the spatial Lagrangian. This leads to the usual Euler--Lagrange equations,
\begin{gather}
\frac{\dd}{\dd T}\left(\frac{\pd \msf{L}}{\pd V^a}\right) 
= \frac{\pd \msf{L}}{\pd X^a}.
\end{gather} 

The spatial Lagrangian can be explicitly written as $\smash{\msf{L} = \msf{L}^{(0)} - \Phi}$, where $\msf{L}^{(0)}_T = - m (\msf{U}^0)^{-1}$ and (cf. \Sec{sec:noncovar0})
\begin{gather}
(\msf{U}^0)^{-1} \doteq \sqrt{-\bar{g}_{00} - 2\bar{g}_{0a} V^a - \bar{g}_{ab} V^a V^b},\\
\Phi \doteq - \bar{L}_\tau^{(2)}/\msf{U}^0.\label{eq:Phi22}
\end{gather}
The corresponding OC canonical momenta are $P_a \doteq \pd \msf{L}/\pd V^a$ and the corresponding OC Hamiltonian is $\msf{H} \doteq P_a V^a - \msf{L}$. Then [cf. \Eqs{eq:hamjac1} and \eq{eq:ham3d}], $\msf{H}(T, \vec{X}, \vec{P}) = - \msf{P}_0(T, \vec{X}, \vec{P}) + O(h^2)$, where $\msf{P}_0$ solves
\begin{gather}\label{eq:mcP0}
H^{(0)}(T, \vec{X}, \msf{P}_0(T, \vec{X}, \vec{P}), \vec{P})=0,
\end{gather}
or explicitly [cf. \Eq{eq:psub00}],
\begin{gather}
\msf{P}_0 (T, \vec{X}, \vec{P}) = - \frac{m \bar{\gamma}}{\sqrt{-\bar{g}^{00}}} - \frac{\bar{g}^{0a}}{\bar{g}^{00}}\,P_a,
\label{eq:psub01}
\end{gather}
where $\bar{\gamma} \doteq \sqrt{1 + \bar{\sigma}^{ab}P_a P_b/m^2}$ and \(\bar{\sigma}^{ab} \doteq \bar{g}^{ab} - \bar{g}^{a0}\bar{g}^{b0}/\bar{g}^{00}\). Using the same theorem \cite[Sec.~40]{book:landau1} as the one used in \Sec{sec:covarOC1}, one finds
\begin{gather}\label{eq:HPhi}
\msf{H}(T, \vec{X}, \vec{P}) = - \msf{P}_0(T, \vec{X}, \vec{P}) + \Phi(T, \vec{X}, \vec{P}),
\end{gather}
and one can adopt $\msf{P}^0 = m\msf{U}^0$ when evaluating $\Phi$, so
\begin{multline}
\Phi =\frac{\varepsilon^{\alpha\beta\gamma\delta}}{2\msf{P}^0}\bigg[
\bar{g}_{\beta\gamma}\TT_{\alpha\delta}
+ \frac{k_\mu k^\mu}{4(k^\lambda P_\lambda)^2}\,\TT_{\alpha\beta} \TT_{\gamma\delta} 
\\ -\frac{\TT_{\alpha\beta}}{k^\lambda P_\lambda}k_\gamma P_\delta
\bigg]_{P_0 = \msf{P}_0}.
\label{eq:ponderpotential1}
\end{multline}
The corresponding Hamilton's equations are
\begin{gather}
\frac{\dd X^a}{\dd T} =\frac{\pd\msf{H}}{\pd P_a},
\quad 
\frac{\dd P_a}{\dd T} =-\frac{\pd\msf{H}}{\pd X^a}.
\label{eq:hamilton1}
\end{gather}

According to \Eq{eq:HPhi}, $-\msf{P}_0$ serves as the free-motion OC Hamiltonian, and \(\Phi\) serves as the interaction Hamiltonian in the OC representation, or the \textit{ponderomotive energy}. Similar terms in electromagnetic wave--particle interactions are often called ponderomotive potentials; however, remember that $\Phi$ depends not only on $T$ and $\vec{X}$ but on $\vec{P}$ too, so it is not a potential \textit{per se} but a more general part of the OC Hamiltonian.

\section{Oscillation-center dynamics: field-theoretical approach}
\label{sec:fieldth}

The calculations above are somewhat \textit{ad hoc} and the final results [\eg \Eq{eq:ponderpotential1}] are not particularly transparent. Here, we propose an alternative derivation of these results that, hopefully, makes them more understandable. The form of the equations derived below will also be advantageous for the discussion in \Sec{sec:polar}.

\subsection{Semiclassical particle model}
\label{sec:klein}

Let us consider a particle as a quantum wave. Since we are not interested in spin effects, we shall assume that this wave is governed by the Klein--Gordon equation,
\begin{gather}
g^{\alpha\beta} \nabla_\alpha \nabla_\beta \psi - m^2 \psi = 0
\end{gather}
(assuming units such that $\hbar = 1$), for it is a simple enough equation that leads to \Eqs{eq:oorays} in the classical limit, as discussed below. Since this equation is linear and has real coefficients, the scalar state function $\psi$ can be assumed real or complex. We choose the latter for simplicity. (The other choice leads to the same final results up to notation.) Then, the corresponding action is $S = \int \mcc{L}\,\dd^4 x$, where $\mcc{L}$ is the Lagrangian density given~by
\begin{gather}
\mcc{L} = \frac{\sqrt{-g}}{2m}
\left(g^{\alpha\beta}\pd_{\alpha} \psi^* \pd_{\beta} \psi
- m^2|\psi|^2\right)
\end{gather}
and $g \doteq \det g_{\alpha\beta}$. Let us represent the wavefunction in the Madelung form, $\smash{\psi = a \ee^{\ii\vartheta}}$ (where $a$ and $\vartheta$ are real), and assume the semiclassical (\ie GO) limit, in which $p \doteq \nabla \vartheta$ is much larger than $\nabla a$. Then, $\mcc{L}$ can be approximated~as
\begin{gather}
\mcc{L} = -\mc{I}(x) H(x, \nabla\vartheta),
\label{eq:lagrdensity}
\end{gather}
where \(\mc{I} \doteq a^2\sqrt{-g}\) and $H$ is given by \Eq{eq:H}. There are two motion equations that flow from here. One is $\delta S[\mc{I}, {\vartheta}]/\delta\mc{I} = 0$, which leads~to
\begin{gather}
H(x, \nabla \vartheta)=0.
\label{eq:euler01}
\end{gather}
This can be recognized as a Hamilton--Jacobi equation \cite{book:landau1}, with $H$ serving as the Hamiltonian; hence, it readily leads to \Eqs{eq:orays} for point particles. The other motion equation is $\delta S[\mc{I}, \vartheta]/\delta \vartheta = 0$, which leads~to
\begin{gather}
\frac{\pd}{\pd x^\alpha}\left[
\mc{I}(x)\,
\frac{\pd H(x, p)}{\pd p_\alpha}
\right]=0.
\label{eq:euler02}
\end{gather}
Equation \eq{eq:euler02} is understood as a continuity equation that represents the action conservation for Klein--Gordon waves, \ie particle conservation. For more details on linear GO as a field theory, see for example, \Refs{book:tracy, my:amc, my:wkin}.

\subsection{Semiclassical OC model}
\label{sec:fieldoc}

Now, let us consider how a semiclassical particle is affected by metric oscillations produced by a GW. To do that, let us represent the Hamiltonian as
\begin{gather}
\label{eq:H012}
H \simeq H^{(0)} + H^{(1)} + H^{(2)}, 
\end{gather}
where $H^{(n)} = O(h^n)$ and higher-order terms are neglected. Then using \Eqs{eq:H} and \eq{eq:hupper}, we find that $H^{(0)}$ is given by \Eq{eq:H0}~and
\begin{gather}
H^{(1)} = -\frac{1}{2m}\,h^{\alpha\beta}p_\alpha p_\beta,
\label{eq:H1}
\\
H^{(2)}=\frac{1}{2m}\,{h^\alpha}_\gamma h^{\gamma\beta}p_\alpha p_\beta.
\label{eq:H2}
\end{gather}
Then, like in \Sec{sec:ponderbasic}, the particle action can be approximated as $S = \int \bar{\mcc{L}}\,\dd^4x$. Here, $\bar{\mcc{L}}= \favr{\mcc{L}}_x$ serves as the Lagrangian density of the slow motion and, under the GO approximation adopted in \Sec{eq:mmodel}, one can also be written as $\bar{\mcc{L}} = \favr{\mcc{L}}_\theta$.

The remaining calculation is similar to that in \Refs{my:lens, my:qponder}, where it was studied how adiabatic propagation of a general linear wave (in our case, a semiclassical particle) is affected by a general quasiperiodic modulation (in our case, a GW) of the general underlying medium (in our case, a background metric). For completeness, we also rederive the corresponding general $\bar{\mcc{L}}$ in \App{app:wave} and show that
\begin{gather}
\label{eq:mcLOC}
\bar{\mcc{L}}=-\bar{\mc{I}}\mc{H}(x, \nabla\bar{\vartheta}),
\end{gather}
where $\bar{\mc{I}} \doteq \favr{\mc{I}}_\theta$, $\bar{\vartheta} \doteq \favr{\vartheta}_\theta$, and
\begin{gather}
\mc{H} = H^{(0)} +\favr{H^{(2)}}_\theta-\frac{k_\mu}{2}\frac{\pd}{\pd P_\mu}
\bigg(
\frac{\favr{{H^{(1)}}^2}_\theta}{k_\lambda \msf{U}^\lambda}
\bigg).
\label{eq:ponderhamiltonian0}
\end{gather}
Here, all $H^{(n)}$ are evaluated on $(x, P)$, $P \doteq \nabla \bar{\vartheta}$, and 
\begin{gather}
\msf{U}^\lambda(X, P) \doteq \frac{\pd H^{(0)}(X, P)}{\pd P_\lambda},
\label{eq:mcU}
\end{gather}
or in our case specifically,
\begin{gather}
\msf{U}^\lambda(X, P) = \bar{g}^{\lambda\alpha}(X)P_\alpha/m.
\end{gather}
The function $\mc{H}$ is introduced here anew but it is, in fact, the same function as in \Eq{eq:mcHc1}. Indeed, let us express it as $\mc{H}=H^{(0)}+\mc{H}^{(2)}$, where $H^{(0)}$ is given by \Eq{eq:H0} and \(\mc{H}^{(2)}\) is inferred from \Eq{eq:ponderhamiltonian0}~to~be
\begin{gather}
\mc{H}^{(2)}=\frac{\varepsilon^{\alpha\beta\gamma\delta}}{2m}\left[
\bar{g}_{\beta\gamma}\TT_{\alpha\delta} 
-
\frac{k_\mu}{4}\frac{\pd}{\pd P_\mu}
\left(
\frac{\TT_{\alpha\beta} \TT_{\gamma\delta}}{k^\lambda P_\lambda}
\right)
\right],
\label{eq:interactionhamiltonian}
\end{gather}
with $\varepsilon^{\alpha\beta\gamma\delta}$ given by \Eq{eq:varepsilon} and $\TT_{\alpha\beta}$ given by \Eq{eq:TT}. A direct calculation shows that \Eq{eq:interactionhamiltonian} is equivalent to \Eq{eq:mcHc3}.

Like in the case of the original system (\Sec{sec:klein}), the corresponding motion equations are as follows:
\begin{gather}
\mc{H}(x, \nabla\bar{\vartheta})=0,
\label{eq:euler1}
\\
\frac{\pd}{\pd x^\alpha}\left[
\bar{\mc{I}}(x)\,
\frac{\pd\mc{H}(x, P)}{\pd P_\alpha}
\right]=0.
\label{eq:euler2}
\end{gather}
Equation \eq{eq:euler1} can be recognized as a Hamilton--Jacobi equation in which $\mc{H}$ serves as a Hamiltonian. Hence, it readily leads to the same Hamilton's equations that we derived earlier, \Eqs{eq:hameqs}. Equation \eq{eq:euler2} is a continuity equation that represents the action conservation of the waves governed by the Lagrangian density \eq{eq:mcLOC}, \ie OC conservation. We shall revisit this equation in \Sec{sec:gravpolariz}.

\subsection{Non-covariant representation}
\label{sec:noncovar}

Since $\mc{H}^{(2)}$ is small, \Eq{eq:euler1} indicates that at a given \((T, \vec{X}, \vec{P})\), the value of \(P_0\) remains close to \(\msf{P}_0(T, \vec{X}, \vec{P})\) that is defined via \Eq{eq:mcP0}. By Taylor-expanding \(\mc{H}(X, P)\) in \(P_0\) around \(\msf{P}_0(T, \vec{X}, \vec{P})\), one obtains
\begin{multline}\label{eq:aux2}
\mc{H}(X, P) = \mc{H}^{(2)}(T, \vec{X}, \msf{P}_0(T, \vec{X}, \vec{P}), \vec{P})  \\
+ \mc{U}^0(T, \vec{X}, \vec{P})[P_0 - \msf{P}_0(T, \vec{X}, \vec{P})],
\end{multline}
where we have introduced [cf. \Eq{eq:mcU}]
\begin{multline}
\mc{U}^0(T, \vec{X}, \vec{P}) \doteq \left[\frac{\pd \mc{H}(X, P)}{\pd P_0}\right]_{P_0 = \msf{P}_0(T, \vec{X}, \vec{P})}
\\= \msf{U}^0(T, \vec{X}, \vec{P}) + O(h^2).
\end{multline}
Up to $O(h^4)$, \Eq{eq:aux2} can also be expressed as
\begin{gather}
\mc{H}(X, P) \simeq \mc{U}^0(T, \vec{X}, \vec{P})[P_0+\msf{H}(T, \vec{X}, \vec{P})],
\label{eq:ponderhamiltonian1}
\end{gather}
where $\msf{H} \doteq -\msf{P}_0 + \Phi$ and
\begin{gather}
\Phi(T, \vec{X}, \vec{P})=\frac{\mc{H}^{(2)}(T, \vec{X}, \msf{P}_0(T, \vec{X}, \vec{P}), \vec{P})}{\msf{U}^0(T, \vec{X}, \vec{P})}.
\end{gather}
This agrees with \Eq{eq:Phi22} [in conjunction with \Eq{eq:mcHc1}] and \Eq{eq:HPhi}. Hamilton's equations corresponding to the approximate Hamiltonian \eq{eq:ponderhamiltonian1} are as follows:
\begin{align}
\frac{\dd T}{\dd \mc{T}} 
& = \frac{\pd \mc{H}}{\pd P_0}
= \mc{U}^0,
\label{eq:hamilton00}
\\
\frac{\dd X^a}{\dd \mc{T}} 
& = \frac{\pd \mc{H}}{\pd P_a}
= \mc{U}^0 \frac{\pd \msf{H}}{\pd P_a},
\label{eq:hamilton01}
\\
\frac{\dd P_0}{\dd \mc{T}} 
& = -\frac{\pd \mc{H}}{\pd T}
= -\mc{U}^0 \frac{\pd \msf{H}}{\pd T},
\label{eq:hamilton02}
\\
\frac{\dd P_a}{\dd \mc{T}} 
& = -\frac{\pd \mc{H}}{\pd X^a}
= -\mc{U}^0 \frac{\pd \msf{H}}{\pd X^a},
\label{eq:hamilton03}
\end{align}
where we used that, according to \Eqs{eq:euler1} and \eq{eq:ponderhamiltonian1},
\begin{gather}
P_0 + \msf{H}(t,\vec{X}, \vec{P})=0.
\end{gather}
Let us substitute \Eq{eq:hamilton00} into \Eqs{eq:hamilton01} and \eq{eq:hamilton03}. Then, one arrives exactly at Hamilton's equations \eq{eq:hamilton1}, with $\msf{H}$ serving as the Hamiltonian of the spatial OC dynamics. Using \Eq{eq:interactionhamiltonian}, one finds that
\begin{gather}
\Phi=\frac{\varepsilon^{\alpha\beta\gamma\delta}}{2\msf{P}^0}\left[
\bar{g}_{\beta\gamma}\TT_{\alpha\delta}
-\frac{k_\mu}{4}\frac{\pd}{\pd P_\mu}
\left(
\frac{\TT_{\alpha\beta} \TT_{\gamma\delta}}{k^\lambda P_\lambda}
\right)
\right]_{P_0 = \msf{P}_0}.
\label{eq:ponderpotential0}
\end{gather}
This formula is in agreement with \Eq{eq:ponderpotential1} that we derived earlier within a different approach.

\subsection{Interaction action}
\label{sec:gravpolariz}

Using $S = \int \bar{\mcc{L}}\,\dd^4x$ [where $x = (t, \vec{x})$], \Eq{eq:mcLOC} for $\bar{\mcc{L}}$, \Eq{eq:ponderhamiltonian1} for $\mc{H}$, and $P_0 = \pd_t \bar{\vartheta}$, one can write
\begin{gather}\label{eq:OCnoncovar}
S = -\int N[\pd_t \bar{\vartheta} +\msf{H}(x, \del\bar{\vartheta})]
\sqrt{-\bar{g}}\, \dd^4 x,
\end{gather}
where $\bar{g} \doteq \det \bar{g}_{\alpha\beta}$, $\del \doteq \pd_\vec{x}$, and $N \doteq \bar{\mc{I}} \, \mc{U}^0/\sqrt{-\bar{g}}$, or explicitly,
\begin{gather}
N(x) \doteq 
\frac{\bar{\mc{I}}(x)\,\mc{U}^0[x, \del\bar{\vartheta}(x)]}{\sqrt{-\bar{g}(x)}}.
\label{eq:n}
\end{gather}
[Note that $x$ in \Eq{eq:OCnoncovar} is a dummy integration variable and can just as well be replaced with $X$.] As flows from \Eq{eq:euler2}, $N$ satisfies a continuity equation,
\begin{gather}
\frac{1}{\sqrt{-\bar{g}}}\,
\frac{\pd(\sqrt{-\bar{g}} N)}{\pd t} 
+ \del \cdot (N \vec{V})=0,
\label{eq:continuity}
\end{gather}
where $\vec{V} \doteq \pd\msf{H}/\pd \vec{P}$ is the OC velocity [cf. \Eq{eq:hamilton1}]. This means that $N$ is the OC density, possibly up to some constant factor $C$. To calculate this factor, let us  consider the point-particle limit, \(N(t,\vec{x}) = C\delta [\vec{x}, \vec{X}(t)]\), where $\delta(\vec{x}', \vec{x}'') \doteq \delta(\vec{x'} - \vec{x''})/\sqrt{-\bar{g}(t, \vec{x}')}$ is the generalized delta function \cite{ref:dewitt52}. Then, one can show \cite{my:qlagr} that $S$ given by \Eq{eq:OCnoncovar} becomes
\begin{gather}
\label{eq:action2}
S = C\int\left[P_a \,\frac{\dd X^a}{\dd T}-\msf{H}(T,\vec{X}, \vec{P})\right]\dd T.
\end{gather}
[This can be viewed as a step towards an alternative derivation of \Eqs{eq:hamilton1}, which readily flow from \Eq{eq:action2}.] By comparing \Eq{eq:action2} with the canonical action of a point object with phase-space coordinates $(\vec{X}, \vec{P})$ \cite{book:landau1}, one finds that $C = 1$.

Let us express the OC action as $S = S^{(0)} + S^{(2)}$, where $S^{(0)}$ is the action of a ``free'' OC and $S^{(2)}$ describes the OC interaction with a GW, which is of the second order in $h$. Specifically, we have
\begin{gather}
S^{(0)} = \int N[\pd_t \bar{\vartheta} + H_0(x, \del\bar{\vartheta})]
\sqrt{-\bar{g}}\,\dd^4 x,
\\
S^{(2)} = -\int N \Phi \sqrt{-\bar{g}}\,\dd^4 x.
\label{eq:intaction}
\end{gather}
It can also be convenient to rewrite $S^{(2)}$ explicitly as a bilinear functional of $h_{\mu\nu}$. To do this, let us rewrite $\Phi$ as
\begin{gather}\label{eq:FA}
\Phi \doteq -\frac{1}{2}\,\varepsilon^{\alpha\beta\gamma\delta}
\mc{A}_{\alpha\beta\gamma\delta}
\equiv
-\frac{1}{2}\favr{h^{\alpha\beta}\mc{A}_{\alpha\beta\gamma\delta} h^{\gamma\delta}}
\end{gather}
and accordingly,
\begin{gather}
S^{(2)} = \frac{1}{2}\int N \varepsilon^{\alpha\beta\gamma\delta} 
\mc{A}_{\alpha\beta\gamma\delta}\sqrt{-\bar{g}}\,\dd^4 x.
\label{eq:S2A}
\end{gather}
The linear coefficient $\mc{A}_{\alpha\beta\gamma\delta}$ that enters these formulas is specific up to any tensor that is anti-symmetric with respect to index permutations $\alpha \leftrightarrow \beta$, $\gamma \leftrightarrow \delta$, or $(\alpha, \beta) \leftrightarrow (\gamma, \delta)$. Let us define $\mc{A}_{\alpha\beta\gamma\delta}$ such that it be symmetric with respect to all these permutations. Then,
\begin{gather}
\label{eq:polar}
\mc{A}_{\alpha\beta\gamma\delta}
= -\frac{1}{4\msf{P}^0}\bigg[
Q_{\alpha\beta\gamma\delta} - 
k_\mu\,\frac{\pd}{\pd P_\mu}
\left(
\frac{\TT_{\alpha\beta} \TT_{\gamma\delta}}{k^\lambda P_\lambda}
\right)
\bigg]_{P_0 = \msf{P}_0},\\
Q_{\alpha\beta\gamma\delta} \doteq 
  (\bar{g}_{\beta\gamma}  \TT_{\alpha\delta}
+ \bar{g}_{\alpha\delta} \TT_{\beta\gamma}
+ \bar{g}_{\alpha\gamma} \TT_{\beta\delta}
+ \bar{g}_{\beta\delta}  \TT_{\alpha\gamma})_{P_0 = \msf{P}_0}.
\end{gather}
The significance of \Eq{eq:S2A} and the physical meaning of $\smash{\mc{A}_{\alpha\beta\gamma\delta}}$ is explained below.

\subsection{Gravitational susceptibility}
\label{sec:polar}

Let us now consider the action $S_\Sigma$ of the \smash{``gas\,+\,spacetime''} system,
\begin{gather}\notag
S_\Sigma = S_{\rm EH} + \sum_n [\smash{S^{(0)}_n + S^{(2)}_n}]
= S_{\rm EH} + S^{(2)}_{\rm gas} + \sum_n S^{(0)}_n.
\end{gather}
Here, $S_{\rm EH}$ is the Einstein--Hilbert action \cite{book:landau2}, the summation index $n$ denotes contributions from individual particles, and $\smash{S^{(2)}_{\rm gas} \doteq  \sum_n S^{(2)}_n}$ is the total interaction action. Using \Eq{eq:S2A}, the latter can also be expressed as
\begin{gather}
S^{(2)}_{\rm gas} = \frac{1}{2}\int \varepsilon^{\alpha\beta\gamma\delta} \mc{X}_{\alpha\beta\gamma\delta}\sqrt{-\bar{g}}\,\dd^4 x,
\\
\mc{X}_{\alpha\beta\gamma\delta} 
\doteq \int \mc{A}_{\alpha\beta\gamma\delta}(x, \vec{P})
F(x, \vec{P})\,\dd\vec{P},
\label{eq:chi0}
\end{gather}
where $F$ is the OC phase-space distribution normalized to the OC density, $\int F(x, \vec{P})\,\dd\vec{P} = N(x)$. This can be used to calculate, both conveniently and systematically, self-consistent metric oscillations in a particle gas from the least-action principle $\delta S_\Sigma = 0$. In particular, equations for $h_{\alpha\beta}$ (equivalent to the linearized Einstein equations) can be derived from $\smash{\delta S_\Sigma/\delta h_{\alpha\beta} = 0}$. Since $\smash{S^{(0)}_n}$ are independent of $h_{\mu\nu}$, one obtains
\begin{gather}\label{eq:einstein}
\frac{\delta}{\delta h_{\alpha\beta}}\left(
S_{\rm EH} + 
\frac{1}{2}\int \varepsilon^{\alpha\beta\gamma\delta} \mc{X}_{\alpha\beta\gamma\delta}\sqrt{-\bar{g}}\,\dd^4 x
\right) = 0.
\end{gather}
Within the linear approximation, the OC distribution~$F$ is a prescribed function. (In plasma theory, such distribution is commonly known as $f_0$.) Then, $\smash{\mc{X}_{\alpha\beta\gamma\delta}}$ is prescribed too, and one readily obtains a self-contained linear equation for $h_{\alpha\beta}$. Such calculations will be presented in a follow-up paper. Related calculations for electromagnetic waves are given, for example, in \Refs{my:itervar, my:bgk, my:acti, my:sharm}. 

Note that $\smash{\mc{X}_{\alpha\beta\gamma\delta}}$ serves in \Eq{eq:einstein} as the \textit{gravitational susceptibility}. Correspondingly, $\smash{\mc{A}_{\alpha\beta\gamma\delta}}$ is the per-particle gravitational susceptibility, or gravitational polarizability. Remarkably, these \textit{linear} response functions emerge from a \textit{nonlinear} (second-order) ponderomotive energy \eq{eq:FA}, in which sense ponderomotive effects are never actually negligible in linear theory. (The fundamental connection between the ponderomotive energy and the linear response function is known as the $K$-$\chi$ theorem \cite{ref:cary77, ref:kaufman87, ref:kentwell87b}; see also \Refs{my:lens, my:qponder, my:nonloc, my:kchi}.) Also note that the gravitational susceptibility can be rewritten as follows:
\begin{gather}\label{eq:Xabcd}
\mc{X}_{\alpha\beta\gamma\delta} 
= 
\int
\left(
\frac{\vec{k}\cdot \pd_\vec{P} F}{\omega - \vec{k} \cdot \vec{V}}\,
\TT_{\alpha\beta} \TT_{\gamma\delta}
+
F J_{\alpha\beta\gamma\delta} 
\right) \frac{\dd\vec{P}}{4(\msf{P}^0)^2},
\\
\label{eq:Jabcd}
J_{\alpha\beta\gamma\delta} \doteq 
\frac{\pd(\TT_{\alpha\beta} \TT_{\gamma\delta})}{\pd P_0}
-
\frac{\bar{g}^{00}}{\msf{P}^0}\,\TT_{\alpha\beta} \TT_{\gamma\delta}
- \msf{P}^0 Q_{\alpha\beta\gamma\delta}.
\end{gather}
(For the derivation and an alternative representation of $J_{\alpha\beta\gamma\delta}$, see \App{app:chi}.) Here, the integrand is evaluated at $P_0 = \msf{P}_0(\vec{P})$ \eq{eq:psub01} and the parametrization $k_\alpha = (-\omega, \vec{k})$ is assumed, as usual.

Finally, note the following. Although we assumed, throughout the paper, that $k_\alpha$ is real and that particles are not resonant to a wave [here, this implies $F(x, \vec{P}) = 0$ where $\omega = \vec{k} \cdot \vec{V}$], our \Eqs{eq:einstein}--\eq{eq:Jabcd} are not actually restricted to this case. Our gravitational susceptibility can be extended to complex $k_\alpha$ via analytic continuation as usual \cite{book:stix}, and resonant particles can be systematically introduced using the formalism from \Ref{my:nonloc} such that the final answer is not affected. For example, \Eq{eq:einstein} correctly describes the kinetic Jeans instability as one of GW modes, as will be shown in a follow-up paper. (An alternative, nonrelativistic approach to the kinetic Jeans instability can be found in \Ref{ref:trigger04}.)

\section{Example: gravitational ponderomotive effects in vacuum}
\label{sec:vacuum}

\subsection{Effective metric}

As a special case, let us consider a linear GW pulse in vacuum. Then, the dispersion relation is \(k_\alpha k^\alpha = 0\), and we also assume the Lorenz gauge \(h^{\alpha\beta}k_\beta = 0\). As seen from \Eq{eq:mcHc3}, $\smash{\mc{H}^{(2)}}$ is simplified then and is given~by
\begin{gather}\label{eq:H2v}
\mc{H}^{(2)} = \frac{\varepsilon^{\alpha\beta\gamma\delta}}{2m}\,
\bar{g}_{\beta\gamma}\TT_{\alpha\delta}.
\end{gather}
[As a reminder, $\varepsilon^{\alpha\beta\gamma\delta}$ is given by \Eq{eq:varepsilon}.] By substituting \Eq{eq:H2v} into \Eq{eq:mcHc1} and using \Eq{eq:H0} for $\smash{H^{(0)}}$, one finds that
\begin{gather}\label{eq:HG}
\mc{H} = \frac{1}{2m}\,(\mcc{G}^{\alpha\beta} P_\alpha P_\beta + m^2),
\\
\mcc{G}^{\alpha\beta} \doteq \bar{g}^{\alpha\beta} + \varepsilon^{\alpha\mu\nu\beta} 
\bar{g}_{\mu\nu}.
\label{eq:mccG}
\end{gather}
Since \(\mcc{G}^{\alpha\beta}\) depends only on $X$ and not on $P$, it can be considered as the effective metric seen by a particle in a GW, or more precisely, the \textit{OC metric}. [In principle, $\mc{H}$ can always be brought to the form \eq{eq:HG}, but in the general case, \(\mcc{G}^{\alpha\beta}\) depends on $P$, in which case it cannot be considered simply as a metric.]

\subsection{Motion equations and conservation laws}

For example, let us assume that our background metric $\bar{g}_{\alpha\beta}$ is the Minkowski metric\footnote{According to the Einstein equations, a linear perturbation $h_{\alpha\beta}$ entails a nonlinear modification of the background metric (for example, see \Ref{ref:harte15}), which then can cause additional memory effects (for example, see \Refs{ref:zhang18, ref:zhang17, ref:flanagan19}). We are not concerned with such effects here or in \Sec{sec:num}, which are intended only to illustrate applications of the OC formalism for prescribed $g_{\alpha\beta}$.} $\eta_{\alpha\beta} = \eta^{\alpha\beta} = \text{diag}\,\lbrace -1, 1, 1, 1\rbrace$ and the perturbation is expressed in the transverse traceless (TT) gauge,
\begin{gather}
h_{\alpha\beta}=
\left(
\begin{array}{cccc}
 0 & 0 & 0 & 0 \\
 0 & h_+ & h_\times & 0 \\
 0 & h_\times & -h_+ & 0 \\
 0 & 0 & 0 & 0 
\end{array}
\right),
\label{eq:hTT}
\end{gather}
where we have assumed that the spatial wavector is parallel to the $x^3$ axis. Along with the vacuum dispersion relation, this implies $k_\alpha = (-\omega, 0 , 0, \omega)$. Also notice that
\begin{gather}
h_{\alpha \gamma} h^\gamma{}_\beta = (h_+^2 + h_-^2)\hat{\eta}_{\alpha\beta},
\end{gather}
where we have introduced the transverse part of the Minkowski metric, $\hat{\eta}_{\alpha\beta} = \hat{\eta}^{\alpha\beta} \doteq \text{diag}\,\lbrace 0, 1, 1, 0\rbrace$. Then,
\begin{gather}
\mcc{G}^{\alpha\beta} = \eta^{\alpha\beta} + q \hat{\eta}^{\alpha\beta},
\quad
q \doteq \favr{h_+^2 + h_\times^2}_\theta.
\end{gather}

Let us also assume that $\omega = \text{const}$ and the GW pulse is one-dimensional, \ie its envelope depends only on $t$ and $x^3$ but not on $x^1$ or $x^2$. In vacuum, such envelope can depend on $x$ only through the wave phase $\theta(x)$. This special case is tractable also without the OC formalism, but the OC formalism makes the solution particularly straightforward. Indeed, in this case, one has
\begin{gather}
\frac{\pd q}{\pd T} = - \frac{\pd q}{\pd X_\parallel} = - \omega q'(\theta)
\end{gather}
and $\vec{P}_\perp$ is conserved. (Here and further, $_\parallel$ denotes components parallel to $\vec{k}$ and $_\perp$ denotes components perpendicular to $\vec{k}$.) Also,  \Eqs{eq:hameqs} yield
\begin{gather}
\frac{\dd \vec{X}_\perp}{\dd \mc{T}} 
= [1+q(\theta)]\,\frac{\vec{P}_\perp}{m},
\quad
\frac{\dd X_\parallel}{\dd \mc{T}} 
= \frac{P_\parallel}{m},
\label{eq:examplehameqs1}
\\
\frac{\dd P_\parallel}{\dd \mc{T}} 
= -\frac{\dd P_0}{\dd \mc{T}} 
= -\frac{\omega}{2m} \, P_\perp^2 q'(\theta).
\label{eq:examplehameqs2}
\end{gather}
Note that \Eq{eq:examplehameqs2} implies
\begin{gather}\label{eq:PPc}
P_0 + P_\parallel = \text{const}.
\end{gather}
Since $\bar{\Omega} \doteq \dd \theta(X)/\dd \mc{T}$ can be written as
\begin{multline}\label{eq:OmPP1}
\bar{\Omega}
= k_\alpha \,\frac{\dd X^\alpha}{\dd \mc{T}}
= \frac{k_\alpha}{m}\,\mcc{G}^{\alpha\beta}P_\beta
\\= \frac{\omega}{m}\,(-\eta^{00}P_0 + \eta^{33}P_\parallel)
= \frac{\omega}{m}\,(P_0 + P_\parallel),
\end{multline}
it also remains constant, according to \Eq{eq:PPc}. Then, \Eq{eq:examplehameqs2} can be integrated, yielding that the parallel momentum $P_\parallel$ is given by
\begin{gather}\label{eq:dPpgain}
P_\parallel = \hat{P}_\parallel - \frac{\omega}{2m\bar{\Omega}} \, P_\perp^2 q(\theta),
\end{gather}
where $\smash{\hat{P}} \doteq P(\mc{T}_0)$ is the initial momentum and $\mc{T}_0$ is the initial moment of time. Also, \Eq{eq:OmPP1} for $\bar{\Omega}$ yields
\begin{gather}
\bar{\Omega} = \omega \hat{\gamma} (-1 + \hat{\beta}_\parallel),
\label{eq:OmPP2}
\end{gather}
where $\hat{\gamma}$ is the initial Lorentz factor and $\hat{\vec{\beta}}$ is the initial velocity normalized to $c$,
\begin{gather}
\hat{\gamma} \doteq - \frac{\hat{P}_0}{m},
\quad
\hat{\vec{\beta}} \doteq \frac{\hat{\vec{P}}}{m \hat{\gamma}}.
\end{gather}
Using \Eq{eq:PPc}, one also finds $\Delta P_0 = -\Delta P_\parallel$. A similar calculation for a charge interacting with a one-dimensional vacuum electromagnetic pulse is discussed in \Ref{my:gev}; see also Ref.~\cite[Sec.~47]{book:landau2}.


\subsection{Secular displacement}

The above equations indicate that a particle in a GW pulse experiences a secular displacement $\Delta \vec{\ell}$ from its unperturbed trajectory,
\begin{align}
\Delta \vec{\ell}_\perp 
& \doteq 
\Delta \vec{X}_\perp - \hat{\vec{P}}_\perp\,\Delta\mc{T}/m,
\\
\Delta \ell_\parallel 
& \doteq 
\Delta X_\parallel - \hat{P}_\parallel\,\Delta\mc{T}/m,
\end{align}
just like a point charge does in an electromagnetic pulse \cite[Sec.~47]{book:landau2}. [The symbols $\Delta$ denote the changes of the corresponding quantities between $\mc{T}_0 \to -\infty$ and $\mc{T} \to + \infty$. Assuming $\omega > 0$, this corresponds to $\theta(\mc{T}_0) \to +\infty$ and $\theta(\mc{T}) \to -\infty$, since in this case $\smash{\bar{\Omega}} < 0$.] From \Eqs{eq:examplehameqs1}, together with \Eqs{eq:dPpgain} for $P_\parallel$ and \eq{eq:OmPP2} for $\bar{\Omega}$, one obtains
\begin{gather}
\Delta \vec{\ell}_\perp = \frac{\mcu{Q}\hat{\vec{\beta}}_\perp}{\omega (1-\hat{\beta}_\parallel)},
\quad
\Delta \ell_\parallel = \frac{\mcu{Q}}{2\omega}
\bigg(
\frac{\hat{\vec{\beta}}_\perp}{1-\hat{\beta}_\parallel}
\bigg)^2.
\label{eq:dL}
\end{gather}
Here, $\mcu{Q}$ is a dimensionless integral proportional to the integral of the GW intensity,
\begin{gather}
\mcu{Q} \doteq \int_{-\infty}^\infty q(\theta) \, \dd \theta \sim q_c \omega \ell_p,
\end{gather}
$q_c$ is the characteristic value of~$q$, and $\ell_p$ is the characteristic length of the GW pulse. Note that a long enough pulse can cause a substantial displacement even at small~$q_c$. Also, $\Delta \ell_\parallel \ge 0$; thus, the gravitational ponderomotive effect displaces a particle away from the GW source. Finally, note that $\Delta \vec{\ell}$ vanishes in the frame where $\smash{\hat{\vec{\beta}}_\perp = 0}$; however, a relative displacement for objects with different $\smash{\hat{\vec{\beta}}}$ is generally nonzero.

\section{Numerical simulations}
\label{sec:num}

\begin{figure}
\includegraphics[width = .46\textwidth]{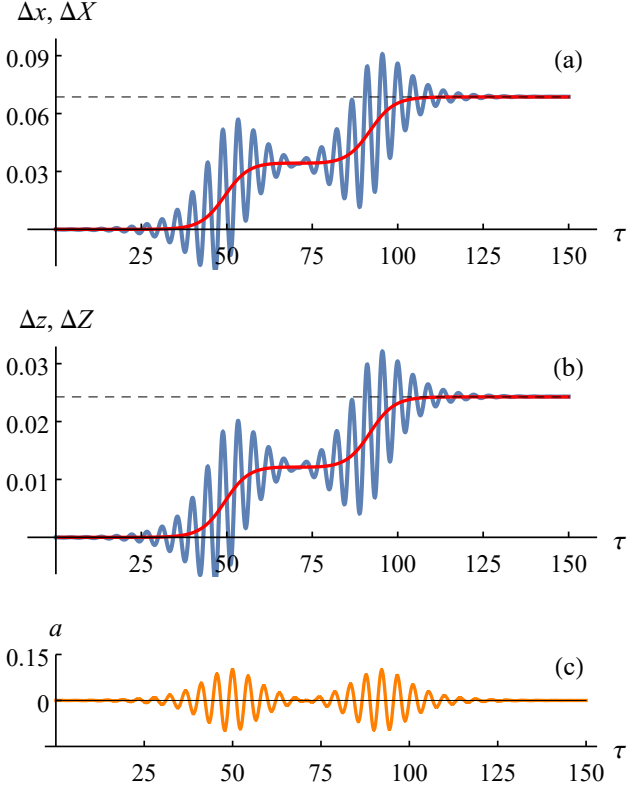}
\caption{Numerical comparison of the particle and OC dynamics in a quasimonochromatic GW: blue -- particle dynamics as predicted by \Eqs{eq:oorays}; red --  the OC dynamics as predicted by \Eqs{eq:hameqs}; black dashed -- $\Delta \ell_x$ and $\Delta \ell_\parallel$ predicted by \Eqs{eq:dL}. The GW propagates along the $z$ axis in vacuum with the Minkowski background metric. Spacetime scales are measured in units $\omega^{-1}$, so the GW wavevector is $k_\alpha = (-1,0,0,1)$. The perturbation metric is given by \Eq{eq:hTT}, with $h_+ = h_\times = a(\theta)/2$, \(a(\theta) = 0.1 [ \sech (\epsilon\theta + 13) - \sech(\epsilon\theta + 7)]\sin{\theta}\), \(\theta = k_\alpha x^\alpha\), and $\epsilon = 0.1$ serves as the small GO parameter (\Sec{eq:mmodel}). The initial velocity is $u^\alpha (\tau=0) = (\sqrt{2},1,0,0)$. Shown are: (a) the transverse displacements relative to the unperturbed trajectory, $\Delta x(\tau)$ and $\Delta X[\mc{T}(\tau)]$; (b) the longitudinal displacements relative to the unperturbed trajectory, $\Delta z(\tau)$ and $\Delta Z[\mc{T}(\tau)]$; (c) the strength of the metric perturbation at the particle location, $a[\theta(\tau)]$. The function $\mc{T}(\tau)$ is calculated by numerical integration of \Eq{eq:Ttau}, but in fact, the difference between $\mc{T}$ and $\tau$ is negligible for these figures.}
\label{fig:vacuum}
\end{figure}

\begin{figure}
\includegraphics[width = .46\textwidth]{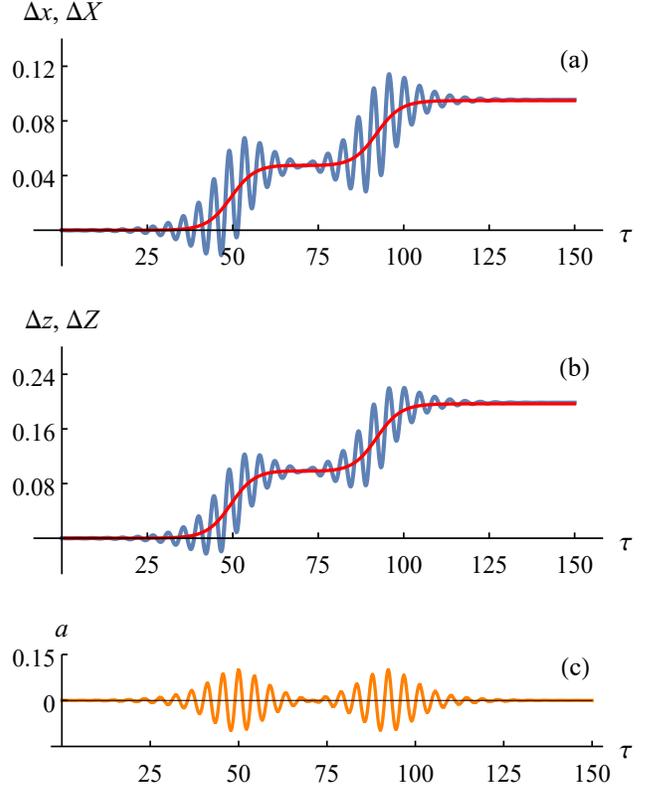}
\caption{Same as \Fig{fig:vacuum} but for non-vacuum dispersion and polarization, namely, $k_\alpha = (-1,0,0,\sqrt{2})$ and $h_{\alpha\beta} =\delta_{\alpha\beta} a(\theta)/2$}.
\label{fig:nonvacuum}
\end{figure}

In order to test our OC theory, we have numerically solved the OC Hamilton's equations [\Eqs{eq:hameqs}] and compared the results with the corresponding numerical solutions of the first-principle equations [\Eqs{eq:oorays}]. Figure~\ref{fig:vacuum} shows the comparison for a linear vacuum GW pulse like those discussed in \Sec{sec:vacuum}. We also compare the total particle displacement $\Delta \vec{\ell}$ from its unperturbed trajectory with the analytic expressions \eq{eq:dL}. Figure~\ref{fig:nonvacuum} shows a similar comparison for an arbitrary non-vacuum GW pulse. (In this case, particle trapping is possible \cite{ref:bruhwiler92, ref:bruhwiler94a}, so there is no general analytic expression for $\Delta \vec{\ell}$ to compare with.) In both cases, the OC theory demonstrates good agreement with first-principle modeling of the particle dynamics. Numerical simulations for other GW profiles, polarizations, wavevectors, and initial conditions have also been done (not shown) and demonstrate good agreement as well.

Finally, as a general comment on test-particle simulations in a prescribed GW, notice the following \cite{book:schutz}. For certain initial conditions and GW polarization, the effect of the wave can be obscured by the coordinate effects in the chosen gauge. For example, the coordinates of a particle that is at rest in the TT gauge remain constant. However, the distance between two such particles can nevertheless change.

\section{Conclusions}
\label{sec:conclusions}

Here, we study the nonlinear secular dynamics of particles in prescribed quasimonochromatic GWs in a general background metric and for general GW dispersion and polarization. We show that this ``ponderomotive'' dynamics can be described by Hamilton's equations \eq{eq:hameqs}, and we derive the corresponding Hamiltonian $\mc{H}$ to the second order in the GW amplitude. We find that $\mc{H} = \smash{H^{(0)}} + \smash{\mc{H}^{(2)}}$, where $\smash{H^{(0)}}$ is given by \Eq{eq:H0} and $\smash{\mc{H}^{(2)}}$ is given by \Eq{eq:mcHc3}, or equivalently, \Eq{eq:interactionhamiltonian}. For the special case of vacuum GWs, we show that our Hamiltonian $\mc{H}$ is equivalent to that of a free particle in an effective metric \eq{eq:mccG}. We also show that already a linear plane GW pulse displaces a particle from its unperturbed trajectory by a finite distance that is independent of the GW phase and proportional to the integral of the pulse intensity. We calculate the particle displacement analytically [\Eq{eq:dL}] and show that our result is in agreement with numerical simulations of the particle motion in a prescribed metric. We also show how the Hamiltonian of the \textit{nonlinear} averaged dynamics naturally leads to the concept of the \textit{linear} gravitational susceptibility of a particle gas with an arbitrary phase-space distribution. This can be understood as a manifestation of the so-called $K$-$\chi$ theorem known from plasma physics. We calculate the gravitational susceptibility explicitly [\Eq{eq:Xabcd}] to apply it, in a follow-up paper, toward studying self-consistent GWs in inhomogeneous media within the geometrical-optics approximation.

This material is based upon the work supported by National Science Foundation under the grant No. PHY~1903130.

\appendix

\section{Field-theoretical calculation of the OC Hamiltonian}
\label{app:wave}

Here, we present a detailed field-theoretical derivation of the general OC Hamiltonian of a semiclassical particle that oscillates in a low-amplitude ``modulating'' wave. The calculation is similar to that in \Ref{my:lens} (see also \Ref{my:qponder}), but the starting point is somewhat different, so we shall restate the whole argument. Suppose a semiclassical particle with quantum phase $\vartheta$ and action density $\mc{I}$. Assume that the particle Lagrangian density $\mcc{L}$ is given by \eq{eq:lagrdensity} and the Hamiltonian $H$ has the form
\begin{gather}
H(x, p) = \bar{H}(x, p) + \widetilde{H}(x, p), 
\\
\bar{H}(x, p) \doteq \favr{H(x, p)}_\theta, \quad
\favr{\widetilde{H}(x, p)}_\theta = 0,
\end{gather}
where $\widetilde{H}(x, p) = \widetilde{H}[\epsilon x, p, \theta(x)]$ is small (cf. \Sec{eq:mmodel}) and the average over the modulating-wave phase $\theta$ is taken at fixed momentum $p \doteq \nabla \vartheta$. (We assume units such that $\hbar = 1$.) Using
\begin{gather}
\vartheta = \bar{\vartheta} + \widetilde{\vartheta},
\quad
\bar{\vartheta} \doteq \favr{\vartheta}_\theta,
\\
\mc{I} = \bar{\mc{I}} + \widetilde{\mc{I}},
\quad
\bar{\mc{I}} \doteq \favr{\mc{I}}_\theta,
\end{gather}
we obtain the following formula for $\mcc{L}$:
\begin{gather}
\mcc{L}
=-(\bar{\mathcal{I}}+\tilde{\mathcal{I}})
[\bar{H} (x, P + \widetilde{p})
+ \widetilde{H}(x, P + \widetilde{p})
],
\end{gather}
where \(P \doteq \nabla \bar{\vartheta}\) and \(\widetilde{p} \doteq \nabla \widetilde{\vartheta}\). Taylor-expanding \(\bar{H}\) and \(\widetilde{H}\) in $\widetilde{p} = O(\smash{\widetilde{H}})$ and neglecting terms of the third and higher orders in $\smash{\widetilde{H}}$, we obtain
\begin{multline}\label{eq:LLLL}
\mcc{L}
\simeq -\bar{\mc{I}} \, \bar{H}
-\bar{\mc{I}} \, \frac{\pd\bar{H}}{\pd P_\alpha} \widetilde{p}_\alpha
-\frac{\bar{\mc{I}}}{2}
\frac{\pd^2\bar{H}}{\pd P_\alpha \pd P_\beta}\, 
\widetilde{p}_\alpha \widetilde{p}_\beta
-\bar{\mc{I}} \, \widetilde{H}
\\
-\bar{\mc{I}} \, \frac{\pd\widetilde{H}}{\pd P_\alpha}\, \widetilde{p}_\alpha
-\widetilde{\mc{I}} \, \bar{H}
-\widetilde{\mc{I}} \, \frac{\pd \bar{H}}{\pd P_\alpha} \widetilde{p}_\alpha
-\widetilde{\mc{I}} \, \widetilde{H},
\end{multline}
where all functions are evaluated at $(x, P)$. From the part of \Eq{eq:euler01} that is linear in the the modulating-wave amplitude, one has
\begin{gather}
\widetilde{H} + \widetilde{p}_\lambda \msf{U}^\lambda = 0,
\quad
\msf{U}^\lambda \doteq \pd\bar{H}/\pd P_\lambda,
\label{eq:eulerquiver}
\end{gather}
so the two last terms on the right-hand side on \Eq{eq:LLLL} mutually cancel out. [The definition of $\msf{U}^\lambda$ given here is in agreement with \Eq{eq:mcU} within the assumed accuracy.] Then, the average Lagrangian density, $\bar{\mcc{L}} \doteq \favr{\mcc{L}}_\theta$, is given by $\bar{\mcc{L}}=-\bar{\mc{I}}\mc{H}$, where
\begin{gather}
\mc{H} = \bar{H}
+
\frac{1}{2}
\frac{\pd^2\bar{H}}{\pd P_\alpha \pd P_\beta}\, 
\langle
\widetilde{p}_\alpha \widetilde{p}_\beta
\rangle_\theta
+
\bigg\langle
\frac{\pd\widetilde{H}}{\pd P_\alpha}\, \widetilde{p}_\alpha
\bigg\rangle_\theta.
\label{eq:avglag0}
\end{gather}
Just like $H$ in \Eq{eq:lagrdensity} serves as a Hamiltonian \textit{for a particle}, $\mc{H}$ serves as a Hamiltonian \textit{for the particle OC}.

The oscillating part of the particle phase is quasiperiodic in $\theta$, so $\smash{\widetilde{\vartheta}} = \smash{\widetilde{\vartheta}}[\epsilon x, \theta(x)]$. Then, $\smash{\widetilde{p}_\alpha} \simeq k_\alpha \pd_\theta\smash{\widetilde{\vartheta}}$, where \(k_\alpha \doteq \nabla_\alpha \theta\) is the wavevector of the modulating wave. Equation \eq{eq:eulerquiver} gives $\pd_\theta\smash{\widetilde{\vartheta}} \simeq -\smash{\widetilde{H}/(k_\lambda \msf{U}^\lambda)}$, so
\begin{gather}
\widetilde{p}_\alpha
\simeq-\frac{k_\alpha \widetilde{H}}{k_\lambda \msf{U}^\lambda}.
\label{eq:ptilde}
\end{gather}
By substituting this into \Eq{eq:avglag0}, we then obtain
\begin{align}
\mc{H}
&\simeq \bar{H}
+
\frac{\pd^2\bar{H}}{\pd P_\alpha \pd P_\beta}\, 
\frac{k_\alpha k_\beta \favr{\widetilde{H}^2}_\theta}{2(k_\lambda \msf{U}^\lambda)^2}
-
\frac{k_\alpha}{k_\lambda \msf{U}^\lambda}
\bigg\langle
\frac{\pd\widetilde{H}}{\pd P_\alpha}\, \widetilde{H}
\bigg\rangle_\theta
\notag\\
& = 
\bar{H}
+
\frac{k_\alpha }{2}\bigg[
k_\beta \,
\frac{\pd \msf{U}^\beta}{\pd P_\alpha}\, 
\frac{\favr{\widetilde{H}^2}_\theta}{(k_\lambda \msf{U}^\lambda)^2}
-
\frac{1}{k_\lambda \msf{U}^\lambda}
\frac{\pd \langle\widetilde{H}^2\rangle_\theta}{\pd P_\alpha}
\bigg]
\notag\\
& = 
\bar{H} - \frac{k_\alpha}{2}
\frac{\pd}{\pd P_\alpha}\bigg( \frac{\favr{\widetilde{H}^2}_\theta}{k_\lambda \msf{U}^\lambda}
\bigg),
\end{align}
where we have used $\pd^2\bar{H}/\pd P_\alpha \pd P_\beta = \pd \msf{U}^\beta/\pd P_\alpha$. For $H$ of the form \eq{eq:H012}, this readily leads to \Eq{eq:ponderhamiltonian0}.

\begin{widetext}

\section{Derivation of the gravitational susceptibility}
\label{app:chi}

Here, we derive an explicit formula for the gravitational susceptibility $\mc{X}_{\alpha\beta\gamma\delta}$ of a particle gas from \Eqs{eq:polar} and \eq{eq:chi0}. By combining the latter equations, one obtains
\begin{gather}
\mc{X}_{\alpha\beta\gamma\delta} 
= -\int \dd\vec{P}\, \frac{F}{4\msf{P}^0} \left[
Q_{\alpha\beta\gamma\delta} 
- 
k_\mu\,\frac{\pd}{\pd P_\mu}
\left(
\frac{\TT_{\alpha\beta} \TT_{\gamma\delta}}{k^\rho P_\rho}
\right)
\right]_{P_0 = \msf{P}_0} 
= (\mc{X}_1 + \mc{X}_2 + \mc{X}_3)_{\alpha\beta\gamma\delta},
\label{eq:Xchi}
\end{gather}
where we have introduced [assuming the parametrization $k_\alpha = (-\omega, \vec{k})$]
\begin{align}
(\mc{X}_1)_{\alpha\beta\gamma\delta}
& \doteq -\int \dd\vec{P}\, \frac{F}{4\msf{P}^0}\,
Q_{\alpha\beta\gamma\delta},
\label{eq:Xchi1}
\\
(\mc{X}_2)_{\alpha\beta\gamma\delta}
& \doteq -\frac{\omega}{4}\int \dd\vec{P}\,\frac{F}{\msf{P}^0}\left[
\frac{\pd}{\pd P_0}
\left(
\frac{\TT_{\alpha\beta} \TT_{\gamma\delta}}{k^\rho P_\rho}
\right)
\right]_{P_0 = \msf{P}_0},
\label{eq:Xchi2}
\\
(\mc{X}_3)_{\alpha\beta\gamma\delta}
& \doteq \frac{k_a}{4}\int \dd\vec{P}\, \frac{F}{\msf{P}^0}\left[
\frac{\pd}{\pd P_a}\left(
\frac{\TT_{\alpha\beta} \TT_{\gamma\delta}}{k^\rho P_\rho}
\right)
\right]_{P_0 = \msf{P}_0}
\notag\\
& = \frac{k_a}{4}\int \dd^4 P\, \delta(P_0 - \msf{P}_0)\, \frac{F}{\msf{P}^0}
\frac{\pd}{\pd P_a}\left(
\frac{\TT_{\alpha\beta} \TT_{\gamma\delta}}{k^\rho P_\rho}
\right).
\end{align}
The latter equality permits taking the corresponding integral by parts. (Remember that the derivative $\pd/\pd P_\mu$ is taken at fixed $P_{\nu \ne \mu}$, which are independent only in the four-dimensional momentum space.) Specifically, one obtains
\begin{align}
(\mc{X}_3)_{\alpha\beta\gamma\delta}
& = -\frac{k_a}{4}\int \dd^4 P\, \delta'(P_0 - \msf{P}_0)
\left(-\frac{\pd \msf{P}_0}{\pd P_a}\right)
\frac{F}{\msf{P}^0}
\frac{\TT_{\alpha\beta} \TT_{\gamma\delta}}{k^\rho P_\rho}
- \frac{k_a}{4}\int \dd^4 P\, \delta(P_0 - \msf{P}_0)
\frac{\pd}{\pd P_a}\left(\frac{F}{\msf{P}^0}\right)
\frac{\TT_{\alpha\beta} \TT_{\gamma\delta}}{k^\rho P_\rho}
\notag\\
& \simeq -\frac{k_a}{4}\int \dd^4 P\, \delta'(P_0 - \msf{P}_0)
\frac{F V^a}{\msf{P}^0}
\frac{\TT_{\alpha\beta} \TT_{\gamma\delta}}{k^\rho P_\rho}
- \frac{k_a}{4}\int \dd\vec{P}\,
\frac{\pd}{\pd P_a}\left(\frac{F}{\msf{P}^0}\right)
\left(
\frac{\TT_{\alpha\beta} \TT_{\gamma\delta}}{k^\rho P_\rho}
\right)_{P_0 = \msf{P}_0}
\notag\\
& = \int \dd^4 P\, \delta(P_0 - \msf{P}_0)\,
\vec{k} \cdot \vec{V}\,\frac{F}{4\msf{P}^0}
\frac{\pd}{\pd P_0}
\left(
\frac{\TT_{\alpha\beta} \TT_{\gamma\delta}}{k^\rho P_\rho}
\right)
- \frac{k_a}{4}\int \frac{\dd\vec{P}}{\msf{P}^0}
\left(
\frac{\pd F}{\pd P_a} - \frac{F}{\msf{P}^0}\frac{\pd P^0}{\pd P_a}
\right)
\left(
\frac{\TT_{\alpha\beta} \TT_{\gamma\delta}}{k^\rho P_\rho}
\right)_{P_0 = \msf{P}_0}
\notag\\
& = \int \frac{\dd\vec{P}}{4\msf{P}^0}
\left[
(\vec{k} \cdot \vec{V})F\,
\frac{\pd}{\pd P_0}
\left(
\frac{\TT_{\alpha\beta} \TT_{\gamma\delta}}{k^\rho P_\rho}
\right)
-
k_a
\left(
\frac{\pd F}{\pd P_a} - \frac{F}{\msf{P}^0}\frac{\pd P^0}{\pd P_a}
\right)
\left(
\frac{\TT_{\alpha\beta} \TT_{\gamma\delta}}{k^\rho P_\rho}
\right)
\right]_{P_0 = \msf{P}_0},
\label{eq:Xchi3}
\end{align}
where we have used $- \pd \msf{P}_0/\pd P_a \simeq \vec{V}$ [see \Eqs{eq:HPhi} and \eq{eq:hamilton1}]. Then, notice that
\begin{gather}
k^\rho P_\rho = k_\rho P^\rho = \msf{P}^0(\vec{k} \cdot \vec{V} - \omega),
\end{gather}
so the sum of \Eqs{eq:Xchi2} and \eq{eq:Xchi3} can be written as follows:
\begin{gather}
(\mc{X}_2 + \mc{X}_3)_{\alpha\beta\gamma\delta}
= \int \frac{\dd\vec{P}}{4\msf{P}^0}
\bigg\{
\frac{F}{\msf{P}^0}
\left[
k^\rho P_\rho\,\frac{\pd}{\pd P_0}
\left(\frac{\TT_{\alpha\beta} \TT_{\gamma\delta}}{k^\rho P_\rho}\right)
+
k^0
\frac{\TT_{\alpha\beta} \TT_{\gamma\delta}}{k^\rho P_\rho}
\right]
-
\left(
k_a\,\frac{\pd F}{\pd P_a} - \frac{F}{\msf{P}^0}\,k_a\frac{\pd P^0}{\pd P_a}
\right)
\left(
\frac{\TT_{\alpha\beta} \TT_{\gamma\delta}}{k^\rho P_\rho}
\right)
\bigg\}_{P_0 = \msf{P}_0}.
\notag
\end{gather}
Notice that $\pd(k^\rho P_\rho)/\pd P_0 = k^0$, so the whole expression in the square brackets is simply $\pd(\TT_{\alpha\beta} \TT_{\gamma\delta})/\pd P_0$. Also,
\begin{gather}
\frac{\pd \msf{P}^0}{\pd P_a} = 
\frac{\pd}{\pd P_a}\,(\bar{g}^{00}\msf{P}_0 + \bar{g}^{0b}P_b)
= \bar{g}^{00}\,\frac{\pd \msf{P}_0}{\pd P_a} + \bar{g}^{0a}
\simeq \bar{g}^{0a} - \bar{g}^{00} V^a,
\\
k_a \,\frac{\pd \msf{P}^0}{\pd P_a}
= \bar{g}^{0a} k_a - \bar{g}^{00} k_a V^a
= k^0 - \bar{g}^{00} k_0 - \bar{g}^{00} k_a V^a
= k^0 + \bar{g}^{00} (\omega - \vec{k} \cdot \vec{V})
= k^0 - \frac{\bar{g}^{00}}{\msf{P}^0}\,k^\rho P_\rho.
\end{gather}
Then, the above equation can be written as follows:
\begin{align}
(\mc{X}_2 + \mc{X}_3)_{\alpha\beta\gamma\delta}
& = \int \frac{\dd\vec{P}}{4\msf{P}^0}
\bigg[
\frac{F}{\msf{P}^0}\,\frac{\pd(\TT_{\alpha\beta} \TT_{\gamma\delta})}{\pd P_0}
-
\left(
k_a\,\frac{\pd F}{\pd P_a} + \bar{g}^{00}k^\rho P_\rho\, \frac{F}{(\msf{P}^0)^2}
\right)
\left(
\frac{\TT_{\alpha\beta} \TT_{\gamma\delta}}{k^\rho P_\rho}
\right)
\bigg]_{P_0 = \msf{P}_0}
\notag\\
& = \int \frac{\dd\vec{P}}{4(\msf{P}^0)^2}\,
\frac{\vec{k}\cdot \pd_\vec{P} F}{\omega - \vec{k} \cdot \vec{V}}\,
(\TT_{\alpha\beta} \TT_{\gamma\delta})_{P_0 = \msf{P}_0}
+
\int \frac{\dd\vec{P}}{4(\msf{P}^0)^2}\,F
\bigg[
\frac{\pd(\TT_{\alpha\beta} \TT_{\gamma\delta})}{\pd P_0}
-
\frac{\bar{g}^{00}}{\msf{P}^0}\,\TT_{\alpha\beta} \TT_{\gamma\delta}
\bigg]_{P_0 = \msf{P}_0}.
\end{align}
Together with \Eqs{eq:Xchi} and \eq{eq:Xchi1}, this leads to
\begin{gather}\label{eq:mcXX}
\mc{X}_{\alpha\beta\gamma\delta} 
= 
\int \frac{\dd\vec{P}}{4(\msf{P}^0)^2}
\left\lbrace
\frac{\vec{k}\cdot \pd_\vec{P} F}{\omega - \vec{k} \cdot \vec{V}}\,
\TT_{\alpha\beta} \TT_{\gamma\delta}
+
FJ_{\alpha\beta\gamma\delta} 
\bigg]
\right\rbrace_{P_0 = \msf{P}_0},
\end{gather}
where
\begin{gather}
J_{\alpha\beta\gamma\delta} 
\doteq
\frac{\pd(\TT_{\alpha\beta} \TT_{\gamma\delta})}{\pd P_0}
-
\frac{\bar{g}^{00}}{\msf{P}^0}\,\TT_{\alpha\beta} \TT_{\gamma\delta}
- \msf{P}^0 Q_{\alpha\beta\gamma\delta}
=
\frac{\pd(\TT_{\alpha\beta} \TT_{\gamma\delta})}{\pd P_0}
- (\msf{P}^0)^2
\left(
\frac{\TT_{\alpha\beta} \TT_{\gamma\delta}}{m^2 + \bar{\sigma}^{ab}P_a P_b}+ Q_{\alpha\beta\gamma\delta}
\right).
\end{gather}
Here, the tensor $\bar{\sigma}^{ab} \doteq \bar{g}^{ab} - \bar{g}^{a0}\bar{g}^{b0}/\bar{g}^{00}$ (same as in \Sec{sec:covarOC2}) is introduced by analogy with $\sigma^{ab}$ in \Eq{eq:tildedh}, and one can further substitute
\begin{gather}
\frac{\pd(\TT_{\alpha\beta} \TT_{\gamma\delta})}{\pd P_0}
=
\delta_\alpha^0 P_\beta P_\gamma P_\delta
+
P_\alpha \delta_\beta^0 P_\gamma P_\delta
+
P_\alpha P_\beta \delta^0_\gamma P_\delta
+
P_\alpha P_\beta P_\gamma \delta^0_\delta.
\end{gather}
\end{widetext}


\end{document}